\documentclass[12pt,preprint]{aastex}

\slugcomment{Submitted to Astrophys. J.}

\shorttitle{Explosive Nucleosynthesis in GRB Jets}
\shortauthors{Nagataki et al.}

\begin{document}


\title{Explosive Nucleosynthesis in GRB Jets \\
    Accompanied by Hypernovae}


\author{Shigehiro Nagataki\altaffilmark{1,2},
Akira Mizuta\altaffilmark{3}, Katsuhiko Sato\altaffilmark{4,5}}

\altaffiltext{1}{Yukawa Institute for Theoretical Physics, Kyoto University,
Oiwake-cho Kitashirakawa Sakyo-ku, Kyoto 606-8502, Japan, E-mail:
nagataki@yukawa.kyoto-u.ac.jp}
\altaffiltext{2}{KIPAC, Stanford University, P.O.Box 20450, MS 29,
Stanford, CA, 94309, USA}
\altaffiltext{3}{Max-Planck-Institute f$\rm \ddot{u}$r Astrophysik, Karl-Schwarzschild-Str.
1, 85741 Garching, Germany}
\altaffiltext{4}{Department of Physics, The University of Tokyo,
Bunkyo-ku, Tokyo 113-0033, Japan}
\altaffiltext{5}{Research Center for the Early Universe (RESCEU),
The University of Tokyo, Tokyo 113-0033, Japan}

\begin{abstract}
Two-dimensional hydrodynamic simulations are performed to investigate
explosive nucleosynthesis in a collapsar using the model of MacFadyen
and Woosley (1999). It is shown that $\rm ^{56}Ni$ is not produced
in the jet of the collapsar sufficiently to explain the observed
amount of a hypernova
when the duration of the explosion is $\sim$10 sec, which is considered
to be the typical timescale of explosion in the collapsar model.
Even though a considerable amount of $\rm ^{56}Ni$ is
synthesized if all explosion energy is deposited initially,
the opening angles of the jets become too wide to realize highly
relativistic outflows and gamma-ray bursts in such a case.
From these results, it is concluded that the origin of $\rm ^{56}Ni$
in hypernovae associated with GRBs is not the explosive
nucleosynthesis in the jet.
We consider that the idea that the origin
is the explosive nucleosynthesis in the accretion disk is more promising. 
We also show that the explosion becomes bi-polar naturally due to the
effect of the deformed progenitor. This fact suggests that the $\rm
^{56}Ni$ synthesized in the accretion disk and conveyed as outflows
are blown along to the rotation axis, which will explain the line
features of SN 1998bw and double peaked line features of SN 2003jd.
Some fraction of the gamma-ray lines from $\rm ^{56}Ni$ decays in the
jet will 
appear without losing their energies because the jet
becomes optically thin before a considerable amount of $\rm ^{56}Ni$
decays as long as the jet is a relativistic flow, which may be
observed as relativistically Lorentz boosted line profiles in future. 
We show that abundance of nuclei whose mass number
$\sim 40$ in the ejecta depends sensitively on the energy deposition
rate, which is a result of incomplete silicon burning and
alpha-rich freezeout. So it may be determined by observations of
chemical composition in metal poor stars which model is the
proper one as a model of a gamma-ray burst accompanied by a hypernova. 
\end{abstract}

\section{INTRODUCTION}\label{intro}

There has been growing evidence linking long gamma-ray bursts (GRBs;
in this study, we consider only long GRBs, so we call long GRBs as
GRBs hereafter for simplicity) to the death of massive stars.
The host galaxies of GRBs are
star-forming galaxies and the position of GRBs appear to trace the
blue light of young stars~\citep{vreeswijk01,bloom02,gorosabel03}.
Also, 'bumps' observed in some afterglows can be naturally explained
as contribution of bright
supernovae~\citep{bloom99,reichart99,galama00,garnavich03}.    
Moreover, direct evidences of some GRBs accompanied by supernovae
have been reported such as the association of GRB 980425 with
SN 1998bw~\citep{galama98,iwamoto98} and that of GRB 030329 with SN 
2003dh~\citep{hjorth03,price03,stanek03}.

It should be noted that these supernovae are categorized as a new
type of supernovae with large kinetic energy ($\sim 10^{52}$ ergs),
nickel mass ($\sim 0.5M_{\odot}$), and
luminosity~\citep{iwamoto98,woosley99}, so these supernovae are
sometimes called
as hypernovae. Also, since GRBs are considered to be jet-like
phenomena~\citep{rhoads99,stanek99}, it is natural to consider
the accompanying supernova to be jet-induced
explosion~\citep{macfadyen99, khokhlov99}.
It is radioactive nuclei, $\rm ^{56}Ni$ and its daughter nuclei,
$\rm ^{56}Co$, that brighten the supernova
remnant and determine its bolometric luminosity. $\rm ^{56}Ni$ is
considered to be synthesized through explosive nucleosynthesis
because its half-life is very short (5.9 days). So it is natural
to consider explosive nucleosynthesis in jet-induced explosion
to understand GRBs accompanied by hypernovae.

Nagataki et al. (1997) have done a numerical calculation of explosive
nucleosynthesis taking account of effects of jet-induced explosion
in the context of normal core-collapse supernova explosion whose 
explosion energy is set to be $10^{51}$ erg. They found that
$\rm ^{56}Ni$ is much produced in the jet region, which means that
much explosive nucleosynthesis occurs around the jet region.
Also, it was found that velocity distribution of iron becomes
double peaked due to the asperical explosion and explosive
nucleosynthesis~\citep{nagataki98,nagataki00}. It was also found that
the velocity distribution, which will be observed as a line profile,
depends on the angle between our line of sight and rotation
axis~\cite{nagataki00}. 
Maeda et al. (2002) have done a numerical calculation of explosive
nucleosynthesis taking account of effects of jet-induced explosion
in the context of hypernovae whose explosion energy is set to be
$10^{52}$ erg. They have shown that sufficient mass of $\rm ^{56}Ni$
enough to explain the observation of hypernovae ($\sim 0.5 M_{\odot}$)
can be synthesized around the jet region
when the explosion energy is set to be $10^{52}$ erg.
They have also calculated line profiles of [$\rm Fe_{\rm II}$]
blend and of [$\rm O_{\rm I}$] from one-dimensional non-LTE nebular
code~\cite{mazzali01} and found that these line profiles depend
on the angle between our line of sight and the direction of the jet.
Recently, it was reported that an asymmetric hypernovae, SN 2003jd
reveal double-peaked profiles in the nebular lines of neutral oxygen
and magnesium~\cite{mazzali05}.

However, here is one question: whether the difference between
explosion energies of $10^{51}$ erg and $10^{52}$ is important or not?
What does the different explosion energy mean? The answer is 'Yes'.
It is impossible to overemphasize its importance because the scenario
of explosion has to be dramatically changed to explain the energetic
explosion energy of $10^{52}$ and to realize a GRB.

Nagataki et al. (1997) investigated explosive nucleosynthesis taking
account of effects of jet-induced explosion in the context of normal
core-collapse supernova explosion, because there is a possibility that
normal core-collapse supernova becomes jet-like when effects of
rotation are taken into account~\citep{yamada94,shimizu94,kotake03}.
In this scenario, the typical timescale of core-collapse supernovae is as
short as $\sim 500$ ms~\cite{wilson85}. So the surrounding layers of
iron core does not collapse so much. As a result, the progenitor
outside the iron core can not be deformed due to the collapse.
This fact supports the
treatment of using a spherical progenitor when explosive nucleosynthesis
is investigated. Note that the central iron core collapses to a
neutron star and it is enough to calculate explosive nucleosynthesis 
in a spherical outer layer such as Si-, O-, and He-rich layers in the
case of normal core-collapse
supernovae~\citep{nagataki97,nagataki98,nagataki98b,nagataki00}. 
Also, to initiate the explosion, the explosion energy is deposited
around the Si-rich layer as an initial condition in their works.
This is justified because the timescale of explosion is very short.

On the other hand, the central engine of GRBs accompanied by
hypernovae is not known well. But it is generally considered that
normal core-collapse supernovae can not cause an energetic explosion
of the order of $10^{52}$ erg. So another scenario has to be
considered to explain the system of GRBs associated with hypernovae.
One of the most promising scenario is the collapsar
scenario~\cite{woosley93}. In the collapsar scenario, a black hole
is formed as a result of gravitational collapse. Also, rotation of
the progenitor plays an essential role. Due to the rotation, an
accretion disk is formed around the equatorial
plane. On the other hand, the matter around the rotation axis falls
into the black hole. It was pointed out that the jet-induced explosion
along to the rotation axis occurs due to the heating through neutrino
anti-neutrino pair annihilation that are emitted from the accretion
disk. MacFadyen and Woosley (1999) demonstrated the numerical
simulations of the collapsar, showing that the jet is launched
$\sim 7$ sec after the gravitational collapse and the duration of the
jet is about 10 sec, which is comparable to the typical
observed duration of GRBs~\citep{mazet81,kouveliotou93,lamb93}. This
timescale is much longer than the typical timescale of
normal core-collapse supernovae.    
As a result, the progenitor becomes deformed even at the Si-rich and
O-rich layer in the collapsar model~\cite{macfadyen99}. In particular,
the density around the rotation axis becomes low because considerable
amount of the matter falls into the black hole, which is a good
environment to produce a fire ball~\citep{woltjer66,rees67}.

Maeda et al. (2002) investigated explosive nucleosynthesis taking
account of effects of jet-induced explosion in the context of
hypernovae whose energy is $10^{52}$ erg using the spherical
progenitor model and depositing explosion energy at the inner
most region initially.  
However, this treatment seems to be incompatible with the collapsar scenario. 
The importance of the duration of explosion, $\sim 10$ sec, is
investigated in some papers~\citep{nagataki03,maeda03} and it was
concluded that the abundance of $\rm ^{56}Ni$ synthesized
during explosion depends sensitively on the duration of explosion
(i.e. energy deposition rate) and $\rm ^{56}Ni$ is not produced
sufficiently to explain the observed amount of $\sim 0.5 M_{\odot}$
when the timescale of explosion becomes as long as 10 sec.
In fact, MacFadyen and Woosley (1999) discussed that enough $\rm
^{56}Ni$ should not be synthesized in the jet in the collapsar model.
Rather, they pointed out the possibility that a substantial amount
of $\rm ^{56}Ni$ is produced in the accretion disk and a part
of it is conveyed outwards by the viscosity-driven
wind~\citep{macfadyen99,pruet02}. There is another question.
Does all of $\rm ^{56}Ni$ produced in the jet of the collapsar model
brighten the supernova remnant? 
If the jet becomes optically thin before $\rm ^{56}Ni$ decays
into $\rm ^{56}Co$ and $\rm ^{56}Co$ decays into $\rm ^{56} Fe$,
these nuclei should result in emitting gamma-rays rather than
brightening the supernova remnant. 

Let us summarize our motivation. We want to understand
how collapsars produce a GRB jet and how collapsars eject
sufficiently enough $\rm ^{56}Ni$ to explain the luminosity
of hypernovae. We want to seek the self-consistent
theory of GRB/Hypernova connection. As a first step, we
want to consider in this work the consistency between the 
collapsar model of MacFadyen and Woosley (1999) and explosive
nucleosynthesis in a hypernova jet. Can a hypernova jet
cause a GRB jet and a sufficiently enough explosive 
nucleosynthesis to explain the luminosity of hypernova?
Rather, should we consider the GRB jet is different from
the hypernova jet? Moreover, should we consider $\rm ^{56}Ni$
comes from the explosive nucleosynthesis in the hypernova jet?
Rather, should we consider $\rm ^{56}Ni$ comes from a different
site? We want to know the answer. This is our motivation of this
work.

Due to the motivation mentioned above. We investigate explosive
nucleosynthesis in the context of the collapsar model.
We use the collapsar model of MacFadyen and Woosley (1999)
in which effects of rotation is included. As a result,
the progenitor becomes deformed significantly as mentioned above.
We show that $\rm ^{56}Ni$ is not produced sufficiently
to explain the observed amount when the duration of the explosion
is $\sim$10 sec, which is consistent with the previous
works~\citep{nagataki03,maeda03}. A fine tuning is required
to explain the amount of $\rm ^{56}Ni$ by the explosive
nucleosynthesis in the jet. 
This result bring us to the conclusion that
the origin of $\rm ^{56}Ni$ in hypernovae associated with GRBs
is not the explosive
nucleosynthesis in the jet but the one in the accretion disk. 
We also show that the explosion
becomes bi-polar naturally due to the effect of deformed
progenitor. This fact suggests that the $\rm ^{56}Ni$ synthesized
in the accretion disk and conveyed as outflows are blown along to
the rotation axis, which can explain the line features of SN 1998bw
and double peaked line features of SN 2003jd~\cite{mazzali05}. 
Also we predict that some fraction of gamma-ray lines from $\rm
^{56}Ni$ decays in the jet may show relativistically Lorentz boosted
line profiles, which might be observed in future.

\section{METHOD OF CALCULATION}\label{method}

We present our method of calculation in this
study. We take account of some effects that had not been included
in Nagataki et al. (2003). In this study, effects of gravitation
and rotation are included. We also adopt realistic equation of state
(EOS) of Blinnikov et al. (1996). Furthermore, we adopt an asymmetric
progenitor model obtained by MacFadyen and Woosley (1999). So we
believe we have done more realistic calculation of explosive
nucleosynthesis in this study compared with Nagataki et al. (2003).

We realize the jet-induced explosion by injecting thermal energy
around the polar region in the same way as MacFadyen and Woosley
(1999) and Aloy et al. (2000). After such a hydrodynamic calculation,
we calculate 
the products of explosive nucleosynthesis as a post-processing.
We explain our detailed method of calculation in the following
subsections.

\subsection{Hydrodynamics}\label{hydro}
\subsubsection{The Scheme}\label{scheme}

We have done two-dimensional hydrodynamic simulations
taking account of self-gravity and gravitational potential of
the central point mass. The calculated
region corresponds to a quarter of the meridian plane under the
assumption of axisymmetry and equatorial symmetry. 
The spherical mesh with 250($r$)$\times$ 30($\theta$) grid points
is used for all the computations. The radial grid is nonuniform,
extending from 2.0$\times 10^{7}$ cm to 3.0$\times 10^{11}$ cm
with finer grids near the center, while the polar grid is uniform.

The basic equations in the following form are finite differenced
on the spherical coordinates:
\begin{eqnarray}
\frac{D \rho}{D t} = && - \rho \nabla \cdot \bf{v} \\
\rho \frac{D \bf{v}}{D t} = && - \nabla p - \rho \nabla \Phi \\
\rho \frac{D}{D t} \left(   \frac{e}{\rho}  \right) = && - p \nabla
\cdot \bf{v},
\end{eqnarray}
where $\rho$, $\bf{v}$, $P$, $\Phi$, and $e$ are density, velocity,
gravitational potential, and internal energy density, respectively.
The Lagrangian derivative is denoted as $D/D t$. The gravitational
potential of the central point mass is modified to account for some of
the effects of general relativity~\cite{paczynski80}, $\phi = - GM/(r
- r_{\rm s})$ where $r_{\rm s} = 2GM/c^2$ is the Schwartzshild radius.
The ZEUS-2D code developed by Stone and Norman (1992) has been used
with an EOS of an electron-positron gas, which is
in thermal equilibrium with blackbody radiation and ideal
gas of nuclei~\cite{blinnikov96}. Since contribution 
of ideal gas of nuclei to the total pressure is negligible relative to
those of electron-positron gas and thermal radiation, we fixed
the mean atomic weight in the EOS to be 16 to calculate total
pressure and temperature using this EOS.

\subsubsection{Initial and Boundary Conditions}\label{initial}

We adopt the 9.15$M_{\odot}$ collapsar model of MacFadyen and
Woosley (1999). When the central black hole has acquired a mass of 3.762
$M_{\odot}$, we map the model to our computational grid. The surface
of the helium star is $R_* = 2.98 \times 10^{10}$ cm. 
Electron fraction, $Y_e$, is set to be 0.5 throughout of this paper
since neutrino process is not included. 

To realize the jet-induced explosion, we deposit only thermal energy
at a rate $\dot{E}=10^{51}$ ergs s$^{-1}$ homogeneously within
a 30$^{\circ}$ cone around the rotation axis for 10 sec.
In the radial direction, the deposition region extends from the
inner grid boundary located at 200km to a radius of 600 km.
This treatment is same as Aloy et al. (2000). We name this model as
Model E51. We consider this model as the standard one.
For comparison, we perform a calculation in which total explosion 
energy (= $10^{52}$ ergs) is put initially with the deposition
region same as Model E51. Also, we perform a calculation
in which total explosion energy is put initially in a spherically symmetric
way (from 200km to 600km). We name these models as Model E52 and E52S,
respectively. We consider that these models represent extreme cases.
Models in this study are summarized in Table~\ref{tab1}.

\placetable{tab1}
 
As for the boundary condition in the radial direction,
we adopt the inflow boundary condition for the inner boundary
while the outflow boundary condition is used for the outer boundary.
That is, the flow toward the central black hole is prohibited
at the inner boundary and the inflow from the surface of the
progenitor is prohibited at the outer boundary. This is because
we consider the phenomenon of explosion, in which the free fall
timescale at the inner boundary will be longer than that of explosion.
It is also noted that we checked that results of explosive
nucleosynthesis
do not depend on the inner boundary condition sensitively by changing the
inflow boundary condition to the outflow boundary condition for the
inner boundary condition. As for the boundary condition in the zenith
angle direction, axix of symmetry boundary condition is adopted for
the rotation axis, while reflecting boundary condition is adopted for
the equatorial plane.

\subsection{Explosive Nucleosynthesis}\label{nucleosynthesis}
\subsubsection{Test Particle Method}\label{test}
Since the hydrodynamics code is Eulerian, we use the test particle
method~\cite{nagataki97}
in order to obtain the informations on the time evolution of the
physical quantities along the fluid motion, which are then used
for the calculations of the explosive nucleosynthesis.
Test particles are scattered in the progenitor and are set
at rest initially. They move with the local fluid
velocity at their own positions after the passage of the shock wave. 
The temperature and density that each test particle
experiences at each time step are preserved.

Calculations of hydrodynamics and explosive nucleosynthesis are
performed separately, since the entropy produced during the explosive
nucleosynthesis is much smaller ($\sim$ a few$\%$) than that generated by
the shock wave.
In calculating the total yields of elements, we
assume that each test particle has its own mass determined
from their initial distribution so that their sum
becomes the mass of the layers where these are scattered.
It is also assumed
that the nucleosynthesis occurs uniformly in each mass element.
These assumptions will be justified since the movement of the test
particles is not chaotic (i.e. the distribution of test particles at
the final
time still reflects the given initial condition) and
the intervals of test particles are sufficiently narrow to give a smooth
distribution of the chemical composition in the ejecta.
The number of the test particles are 1500. 
The test particles are put non-uniformly in the radial direction,
extending from 2.0$\times 10^{7}$ cm to 3.0$\times 10^{10}$ cm with
closely separated near the center, while they are put uniformly in
the polar direction.

\subsubsection{Post-Processing}\label{reaction}
Since the chemical composition behind the shock wave is not in nuclear 
statistical equilibrium, the explosive nucleosynthesis has to be
calculated using the time evolution of $(\rho,T)$ and a nuclear
reaction network, which is called post-processing. We use the data of
$(\rho,T)$ comoving with the matter obtained by the test particle
method mentioned in subsection~\ref{test}. The nuclear reaction network
contains 250 species (see Table~\ref{tabnucl}). We add some species around
$\rm ^{44}Ti$ to Hashimoto's network that contains 242
nuclei~\cite{hashimoto89}, although it turned out that the result was not
changed essentially by the addition.

\placetable{tabnucl}

\section{RESULTS}\label{result}

\begin{figure}
\epsscale{.80}
\plotone{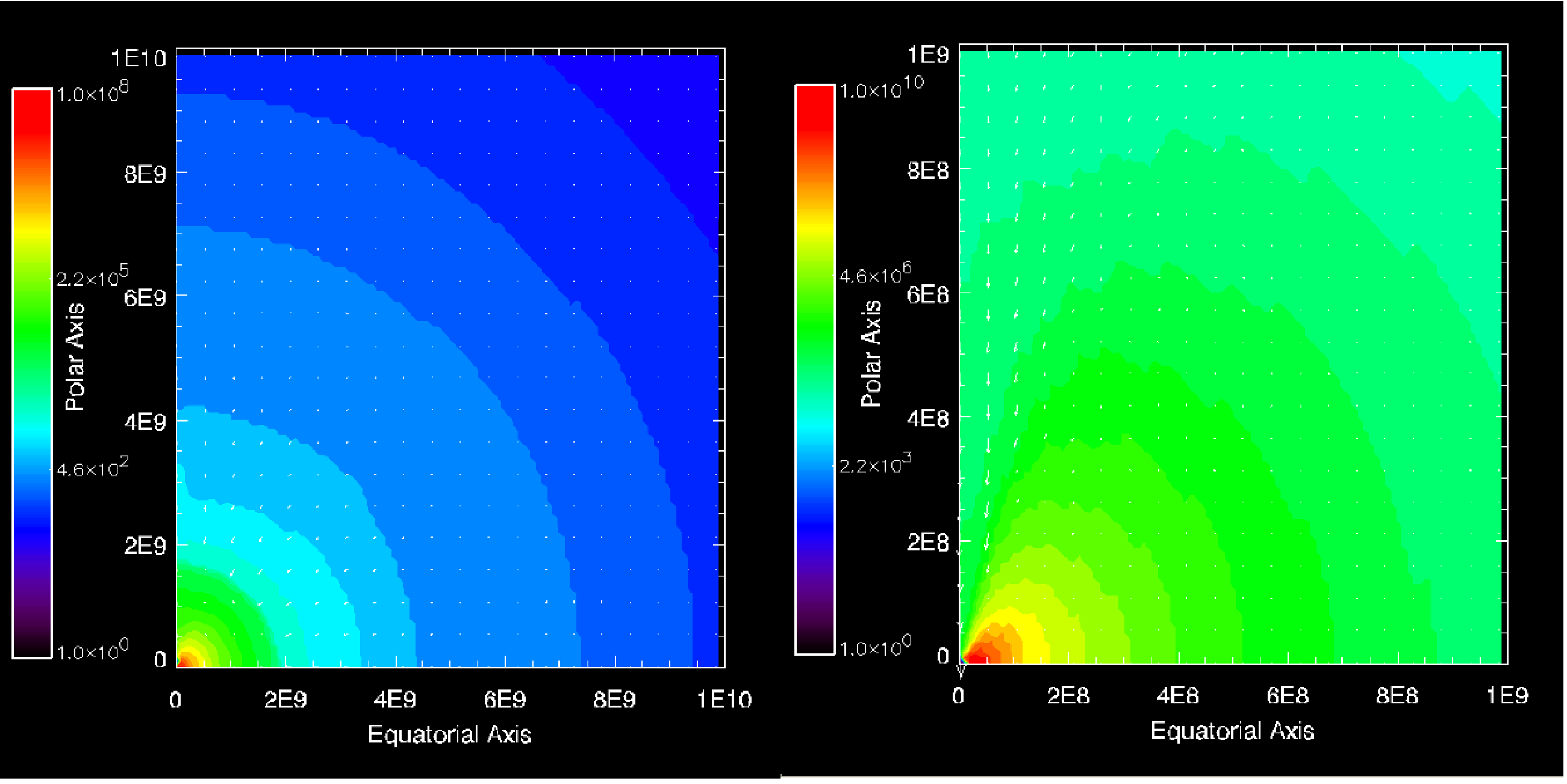}
\caption{Initial density structure in our simulation. This model
is the 9.15$M_{\odot}$ collapsar model from MacFadyen and
Woosley (1999). The mass of the central black hole is 3.762 $M_{\odot}$.
The surface of this helium star is $R_* = 2.98 \times 10^{10}$ cm. 
The color represents the density (g cm$^{-3}$) in logarithmic scale.
The polar axis represents the rotation axis, while the horizontal
axis represents the equatorial plane. The axis is written in
units of cm. The arrow represents the velocity field in ($r$,$\theta$)
plane. The region within $10^{10}$ cm is shown in the left panel,
while that within $10^9$ cm is shown in the right panel.
 \label{fig1}}
\end{figure}

First, initial density structure in our simulation is shown in
Fig.~\ref{fig1}. This model is the 9.15$M_{\odot}$ collapsar model from
MacFadyen and Woosley (1999). The mass of the central black hole is 3.762
$M_{\odot}$~\cite{aloy00}. The surface of this helium star is
$R_* = 2.98 \times 10^{10}$ cm. The color represents the density
(g cm$^{-3}$) in logarithmic scale. The polar axis represents the
rotation axis, while the horizontal axis represents the equatorial
plane. The axis is written in units of cm.
The arrows represent the velocity field in ($r$,$\theta$) plane.
The region within $10^{10}$ cm is shown in the left panel,
while that within $10^9$ cm is shown in the right panel.
An accretion disk is clearly seen in the right panel.
The typical specific angular momentum is
$\sim 10^{17}$ cm s$^{-1}$~\cite{macfadyen99},
although this is not shown in Fig.~\ref{fig1}.

\placefigure{fig1}

\begin{figure}
\epsscale{1.0}
\plotone{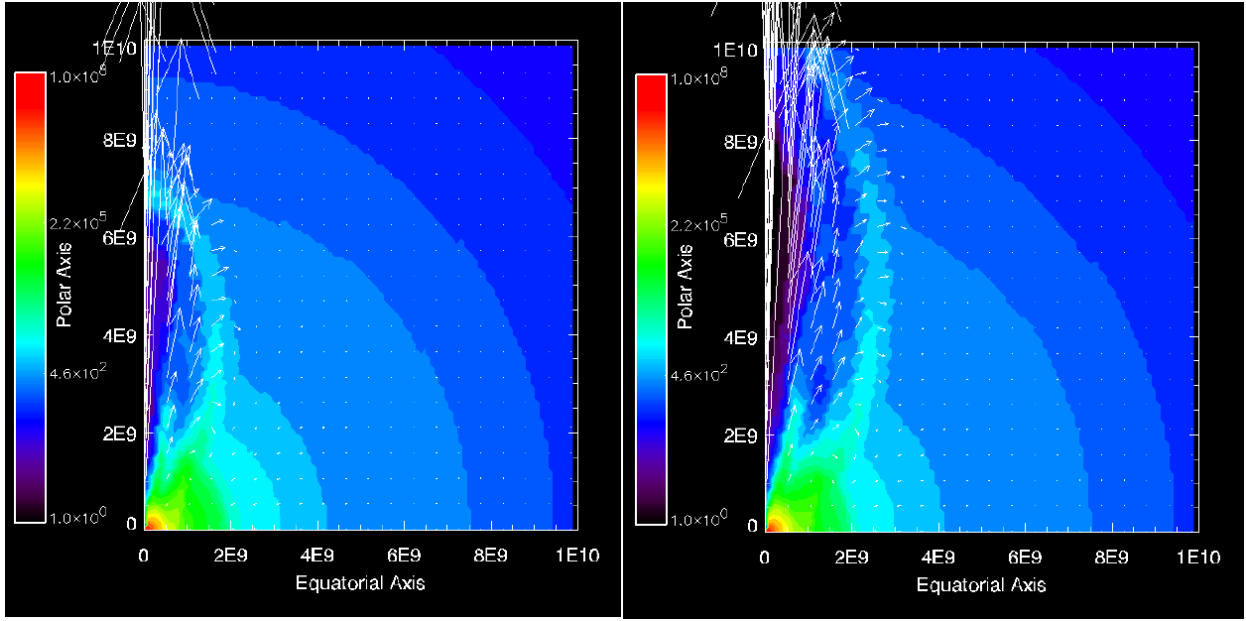}
\caption{Density structure and velocity field of Model E51 at t = 1.0
sec (left panel) and t = 1.5 sec (right panel). 
\label{fig2}}
\end{figure}

\begin{figure}
\epsscale{1.0}
\plotone{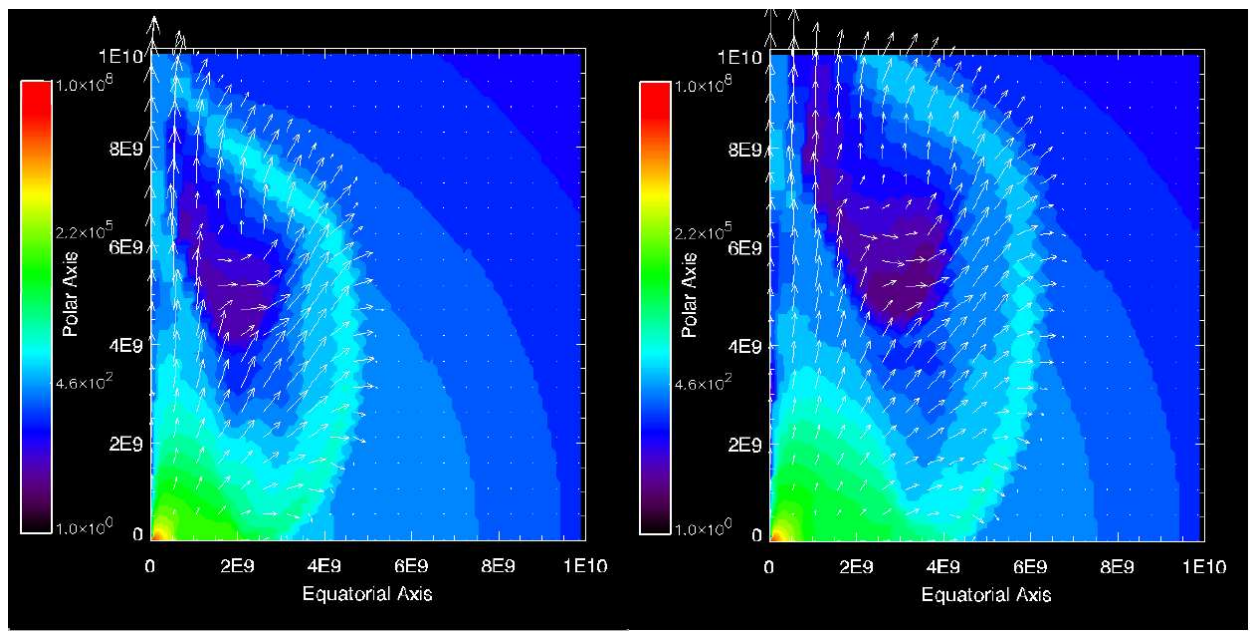}
\caption{Same with Fig.~\ref{fig2}, but for the Model E52 at t = 1.0
sec (left panel) and t = 1.5 sec (right panel). 
 \label{fig3}}
\end{figure}

\begin{figure}
\epsscale{1.0}
\plotone{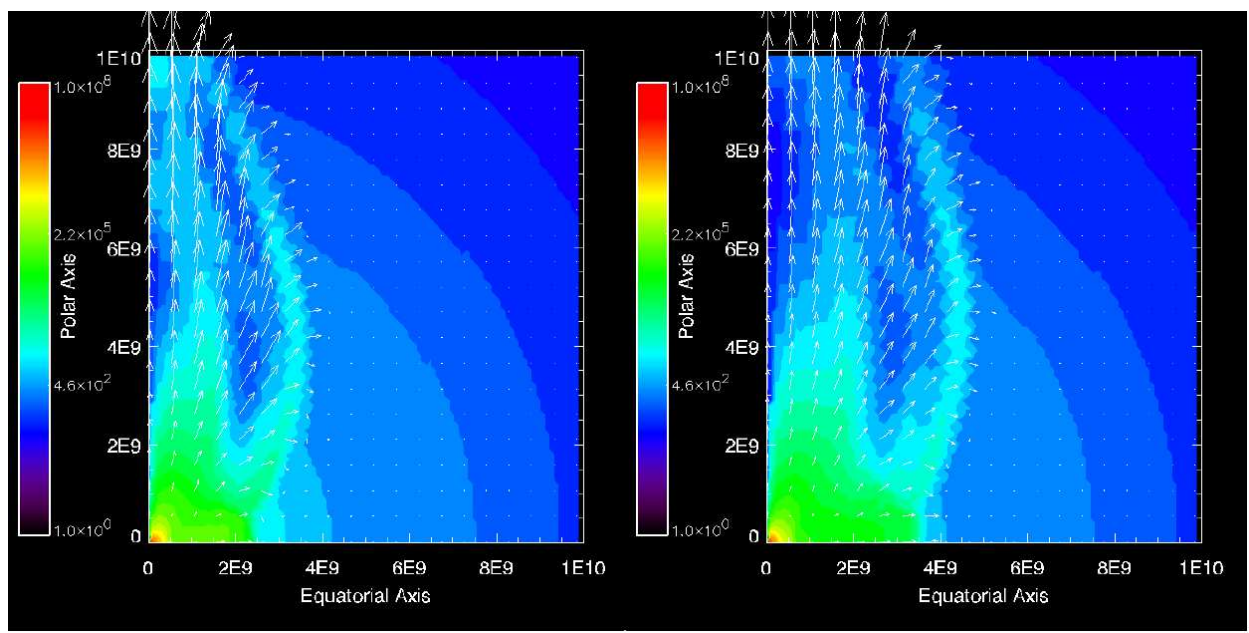}
\caption{Same with Fig.~\ref{fig2}, but for the Model E52s at t = 1.0
sec (left panel) and t = 1.5 sec (right panel). It is clearly shown
that the shock wave is deformed even though the thermal energy is
deposited in a spherically symmetric way as an initial condition. This is due
to the asymmetry of the density structure of the progenitor. 
 \label{fig4}}
\end{figure}

As explained in section~\ref{initial}, we deposit thermal energy
to launch a jet from the central region of the collapsar. The density
structure for Models E51, E52, and E52s at $t=1.0$ sec (left panel)
and $t=1.5$ sec (right panel) are shown in Figs.~\ref{fig2}, \ref{fig3},
and \ref{fig4}. It is clearly shown that a sharp, narrow jet propagates
along to the rotation axis in Model E51, which is similar to Aloy et
al. (2000). On the other hand, in the case of E52, a broad, deformed
shock wave propagates in the progenitor. Also, in Model E52S, the
shock wave is deformed even though the thermal energy is deposited
in a spherically symmetric way. This is due to the asymmetry of the density
structure of the progenitor. 
That is, in the low density region around the rotation axis,
the injected thermal energy is mainly shared with electrons,
positrons, and photons. On the other hand, at the high density
region around the equatorial plane, the injected energy is
shared with radiations mentioned above and nuclei/nucleons.
As a result, the pressure gradient at the energy-injected region
becomes aspherical, which causes a bi-polar flow along to the
rotation axis.
It is also noted that the shock wave is more deformed in 
Model E52S than Model E52. This will be because the energy
density around polar region is higher in Model E52, making 
this region expands very strongly.

\placefigure{fig2}
\placefigure{fig3}
\placefigure{fig4}

\begin{figure}
\epsscale{1.0}
\plotone{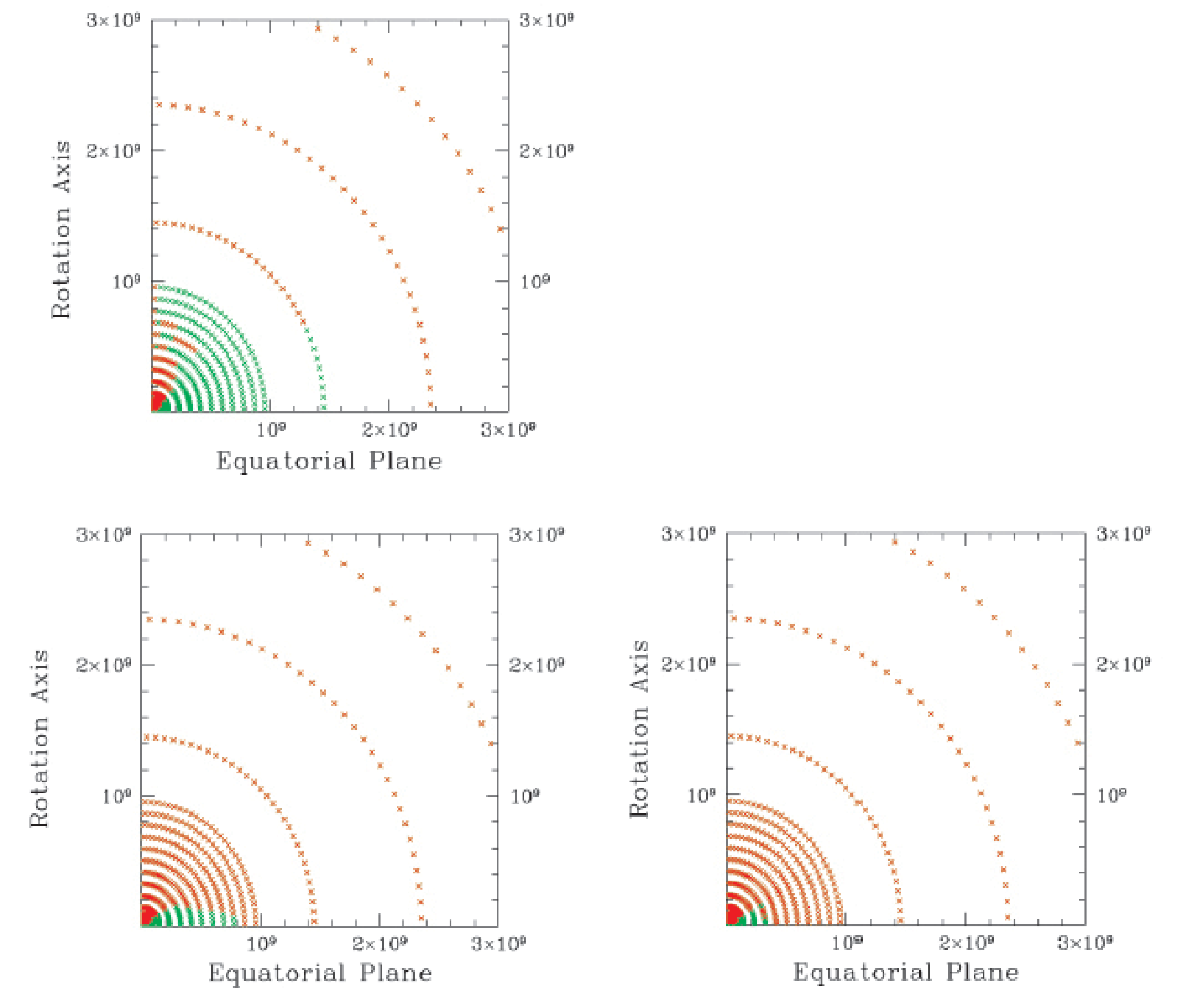}
\caption{Mass cut for the Model E51 (upper panel),
Model E52 (lower left panel), and
Model E52S (lower right panel). 
Red particles represent the ones
that can escape from the gravitational potential to infinity, while
green particles represent the ones that are trapped into the
gravitational potential. 
 \label{fig5}}
\end{figure}

Forms of the mass cut (the boundary between ejecta and the matter that
falls into the central black hole) are shown in Fig.~\ref{fig5}
for each model. Red particles represent the ones that
can escape from the gravitational potential to infinity, while green
particles represent the ones that are trapped into the gravitational
potential. For a criterion to judge whether a test particles can
escape or not, we calculated the total energy (summation of
kinetic energy, thermal energy, and gravitational energy) of the test
particles at the final stage of simulations ($t=10$ sec). We judge
that a test particle can escape if its total energy is positive, and 
vice versa. 

\placefigure{fig5}

Strictly speaking, we have to simulate much longer time
to determine whether a test particle can really escape or not. In
particular, there is a region around the equatorial plane ($r \ge
10^{10}$ cm, $\theta \ge 70^{\circ}$) where shock wave does not reach
even at the final stage of the simulations ($t=10$ sec), because the speed of
propagation of the shock wave is slower around the equatorial plane
compared with the polar region. In this study, we assumed that such a
region where the shock wave does not reach even at the final stage can
escape to infinity. This is because such a region is distant from the
central black hole ($r \ge 10^{10}$ cm), so gravitational potential is
shallow. Moreover, as shown later, at such a distant region, explosive
nucleosynthesis hardly occurs and the most important nuclei in this
study, $\rm ^{56}Ni$, is not synthesized. So our results on the
abundance of $\rm ^{56}Ni$ do not depend on this assumption.

\begin{figure}
\epsscale{1.0}
\plotone{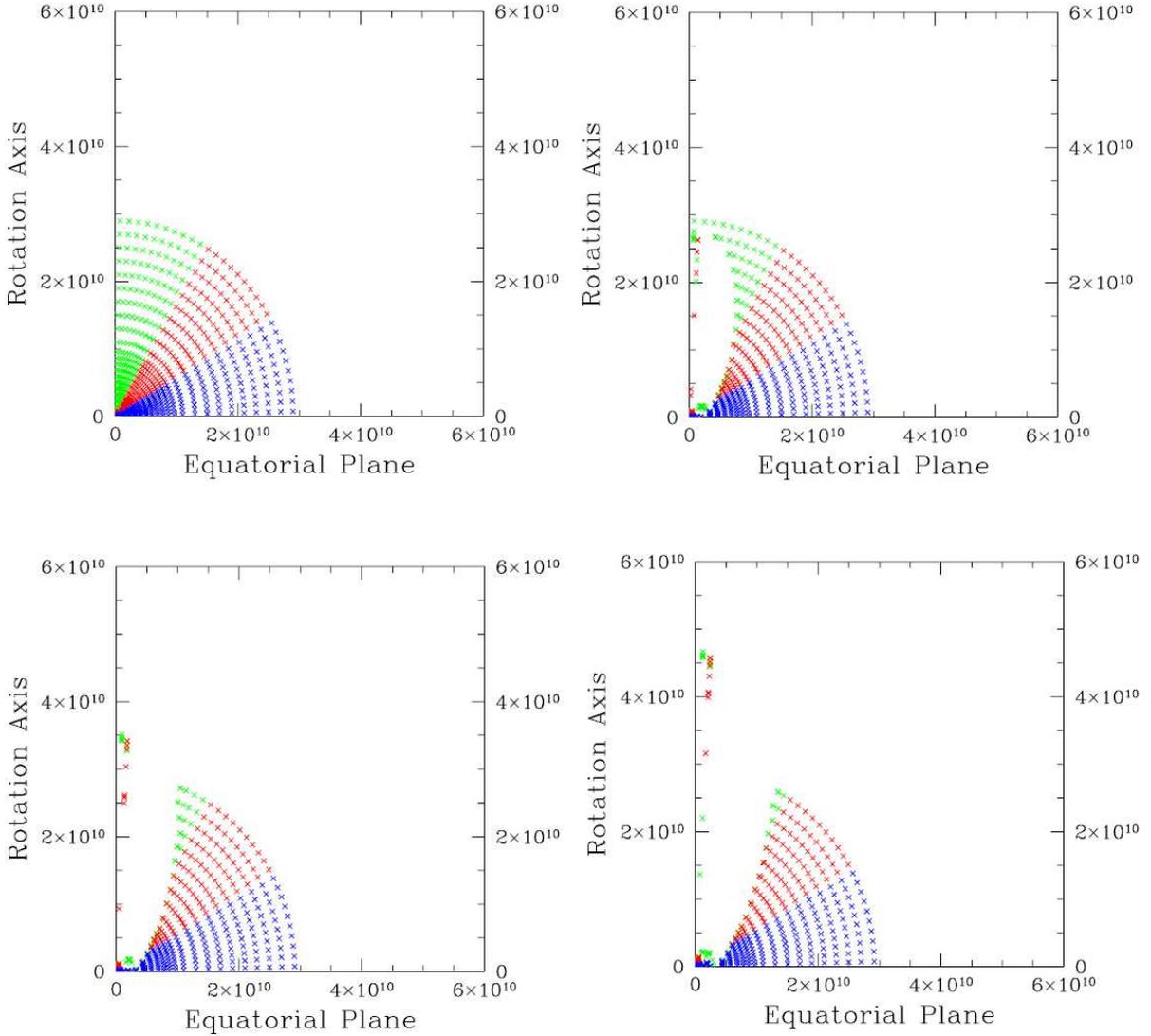}
\caption{
Positions of test particles at $t = 0$ sec (upper left panel),
$t = 3.11 $ sec (upper right panel), 
$t = 3.69 $ sec (lower left panel), and
$t = 4.27 $ sec (lower right panel)
for Model E51. The particles colored
green, red, and blue are the ones that are put around the rotation
axis initially ($\theta \le 30^{\circ}$), middle range ($30^{\circ}
\le \theta \le 60^{\circ}$), and equatorial plane ($60^{\circ}
\le \theta \le 90^{\circ}$), respectively, where $\theta$ is the
zenith angle. Radius of the progenitor is $2.98 \times 10^{10}$ cm.
The (white) region where no test particle exist shows the shocked,
low density region. It is clearly shown that some fraction of
the matter behind the shock wave composes the jet component around
the rotation axis, while some fraction of it is pushed away toward
the $\theta$-direction that composes supernova component.
 \label{fig6}}
\end{figure}

To show how test particles are ejected clearly, we show the 
positions of the test particles at $t = 0$ sec (upper left
panel of Fig.~\ref{fig6}),
3.11 sec (upper right panel of Fig.~\ref{fig6}), 3.69 sec (lower left panel of
Fig.~\ref{fig6}), and 4.27 sec (lower right panel of Fig.~\ref{fig6}) for
Model E51. The particles colored green, red, and blue are the ones
that are put around the rotation axis initially ($\theta \le
30^{\circ}$), middle range ($30^{\circ} \le \theta \le 60^{\circ}$),
and equatorial plane ($60^{\circ} \le \theta \le 90^{\circ}$),
respectively. Radius of the progenitor is $2.98 \times 10^{10}$ cm.
The (white) region where no test particle exist shows the shocked,
low density region. It is clearly shown that some fraction of
the matter behind the shock wave composes the jet component around
the rotation axis, while some fraction of it is pushed sideways toward
the $\theta$-direction, which we call as supernova (SN) component
in this study. We define the jet component as the matter within the
$10^{\circ}$ cone around the rotation axis at the final stage of the
calculation, and the rest we define as SN component. The motions of
test particles in Models E52 and E52S are similar to Model E51.  

\begin{figure}
\epsscale{1.0}
\plottwo{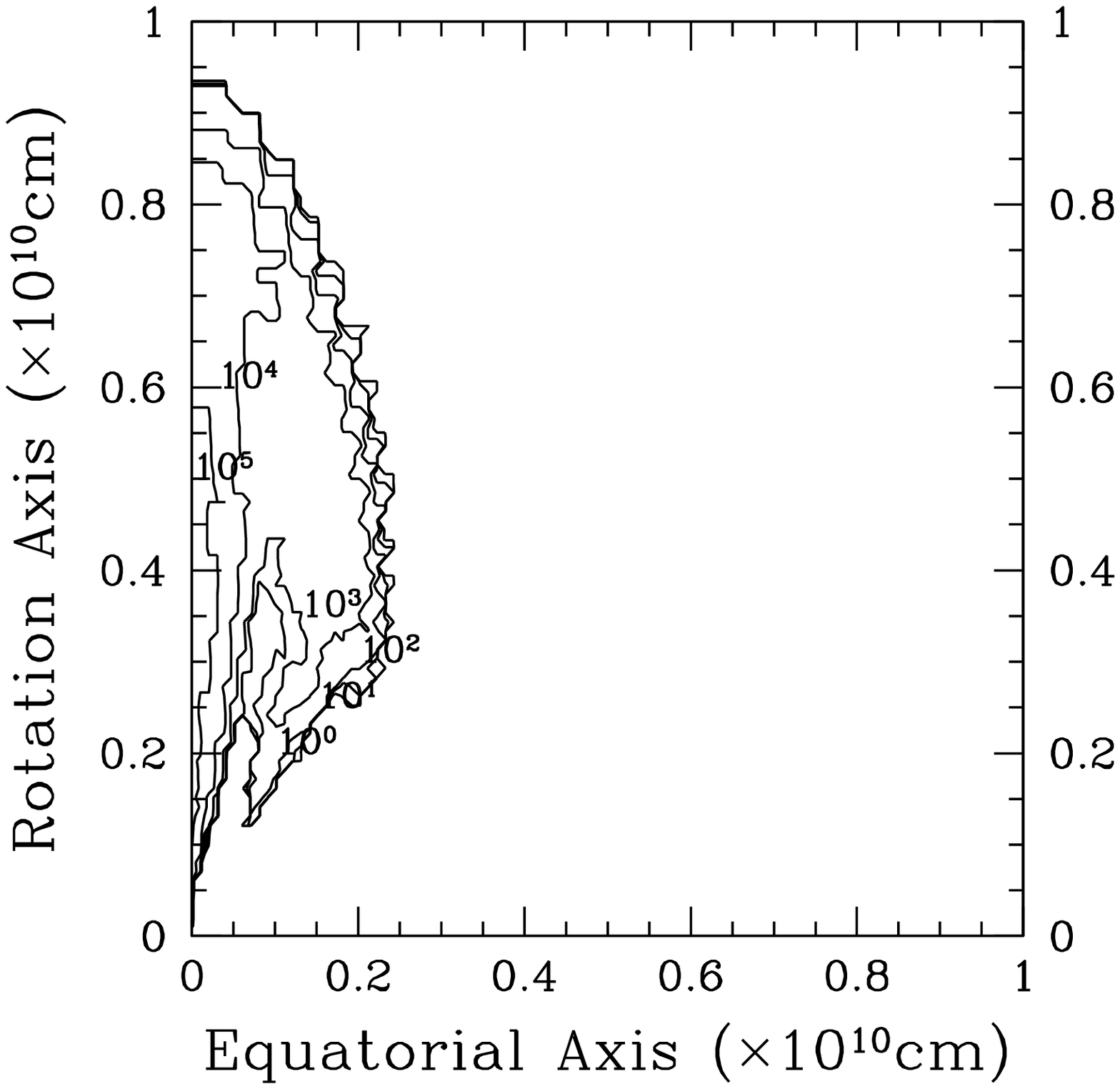}{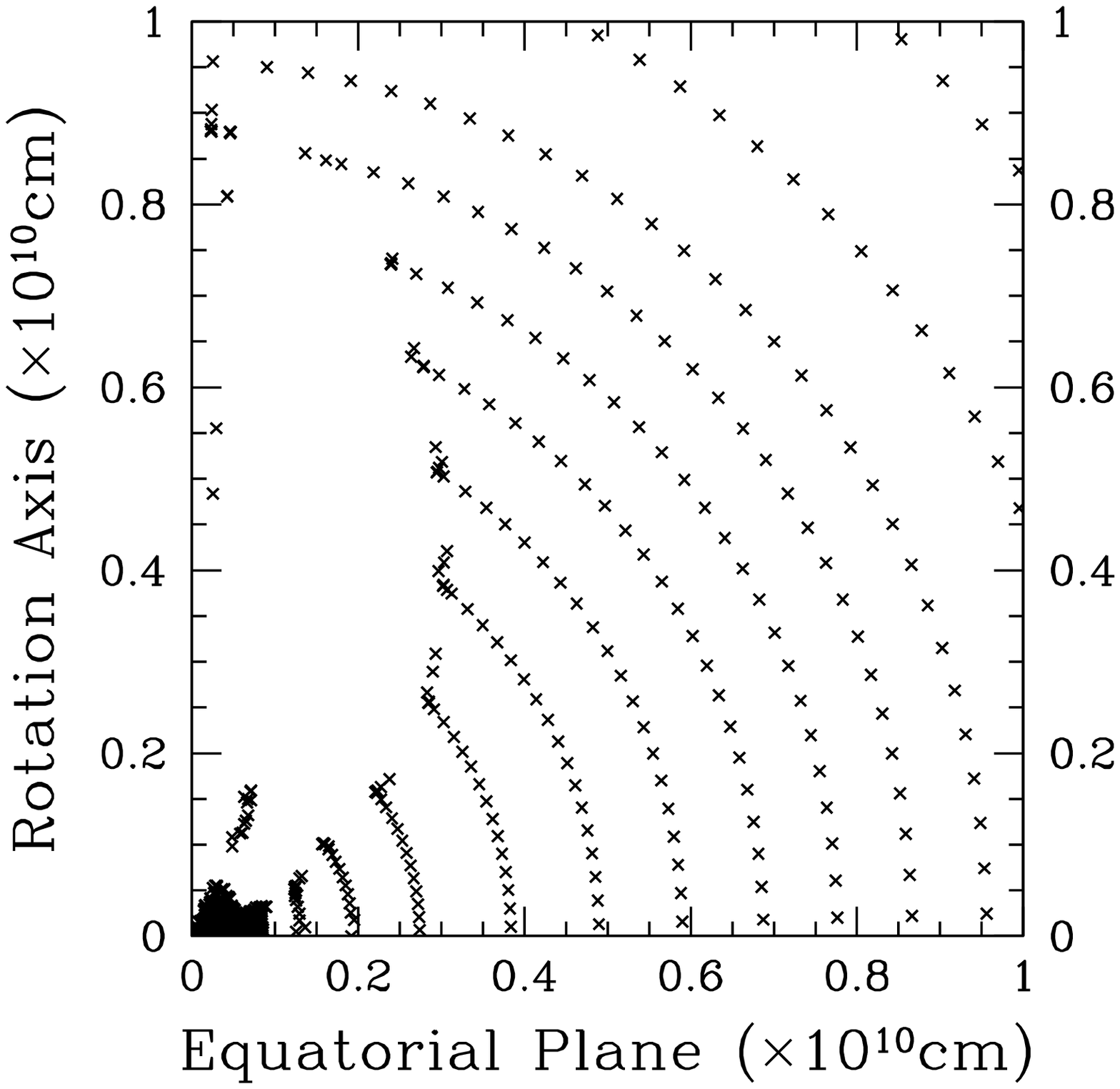}
\caption{
Left panel: contour of entropy per baryon in units of $k_b$ for
Model E51 at $t$ = 1.5 sec. Right panel: positions of test particles
for Model E51 at $t=$ 1.5 sec. As for the entropy per baryon,
the range $10^0-10^5$, which corresponds to the shocked region,
is shown. It is clearly shown that test particles in the downstream
of the shock wave are moving with shock velocity, and no test
particles are left inside of the shocked region where entropy per
baryon is quite high. 
 \label{fig7}}
\end{figure}

Before we show the results of explosive nucleosynthesis, we present
contour of entropy per baryon in units of Boltzmann constant ($k_b$)
for Model E51 at $t$ = 1.5 sec in the left panel of Fig.~\ref{fig7}.  
For comparison, positions of test particles at that time is shown in
the right panel of Fig.~\ref{fig7}. As for the entropy per baryon,
the range $10^0-10^5$, which corresponds to the shocked region,
is shown. It is clearly shown that test particles in the downstream
of the shock wave are moving with shock velocity, and no test
particles are left inside of the shocked region where entropy per
baryon is quite high.

\placefigure{fig6}
\placefigure{fig7}

\begin{figure}
\epsscale{1.0}
\plottwo{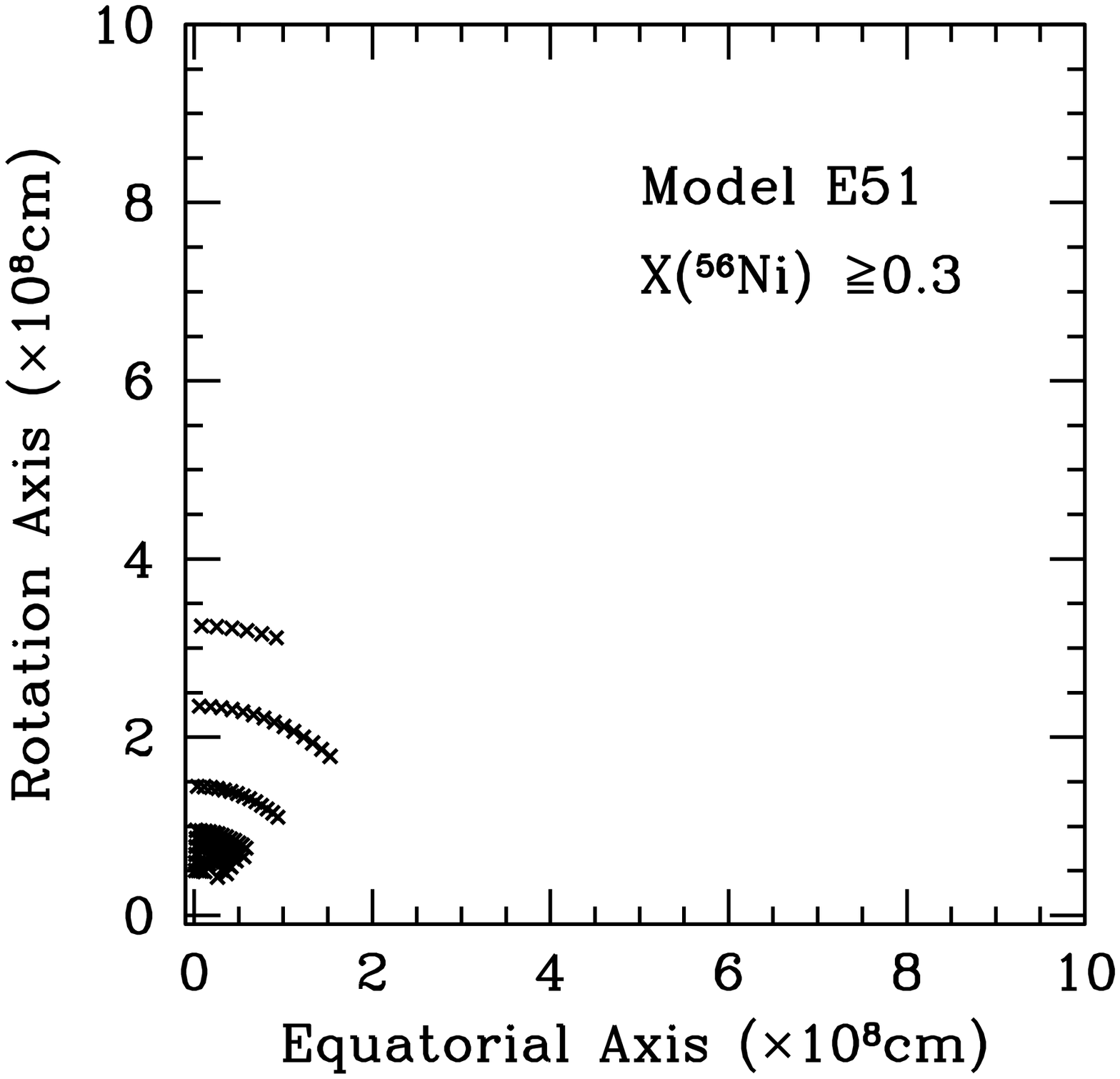}{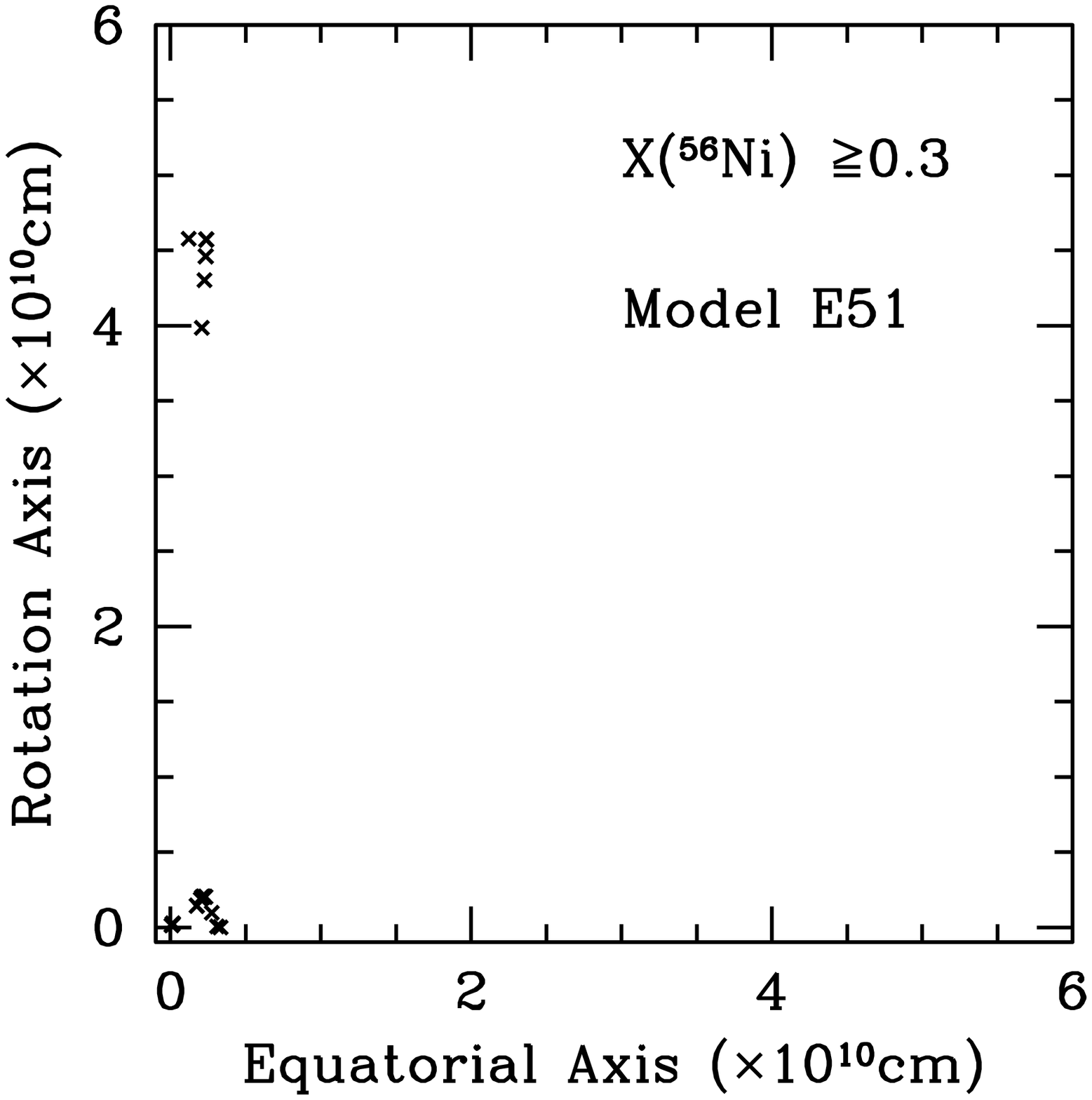}
\caption{Positions of the ejected test particles 
at $t=0$ sec (left panel) and $t= 4.27$ sec (right panel)
that meet the
condition that the mass fraction of $\rm ^{56}Ni$ becomes
greater than 0.3 as a result of explosive nucleosynthesis
for Model E51 . Total ejected mass of $\rm ^{56}Ni$
becomes 0.0439$M_{\odot}$. In particular, total mass of it in the
supernova component is 0.0175$M_{\odot}$, which is much
smaller than the observed value of hypernovae.
 \label{fig8}}
\end{figure}

\begin{figure}
\epsscale{1.0}
\plottwo{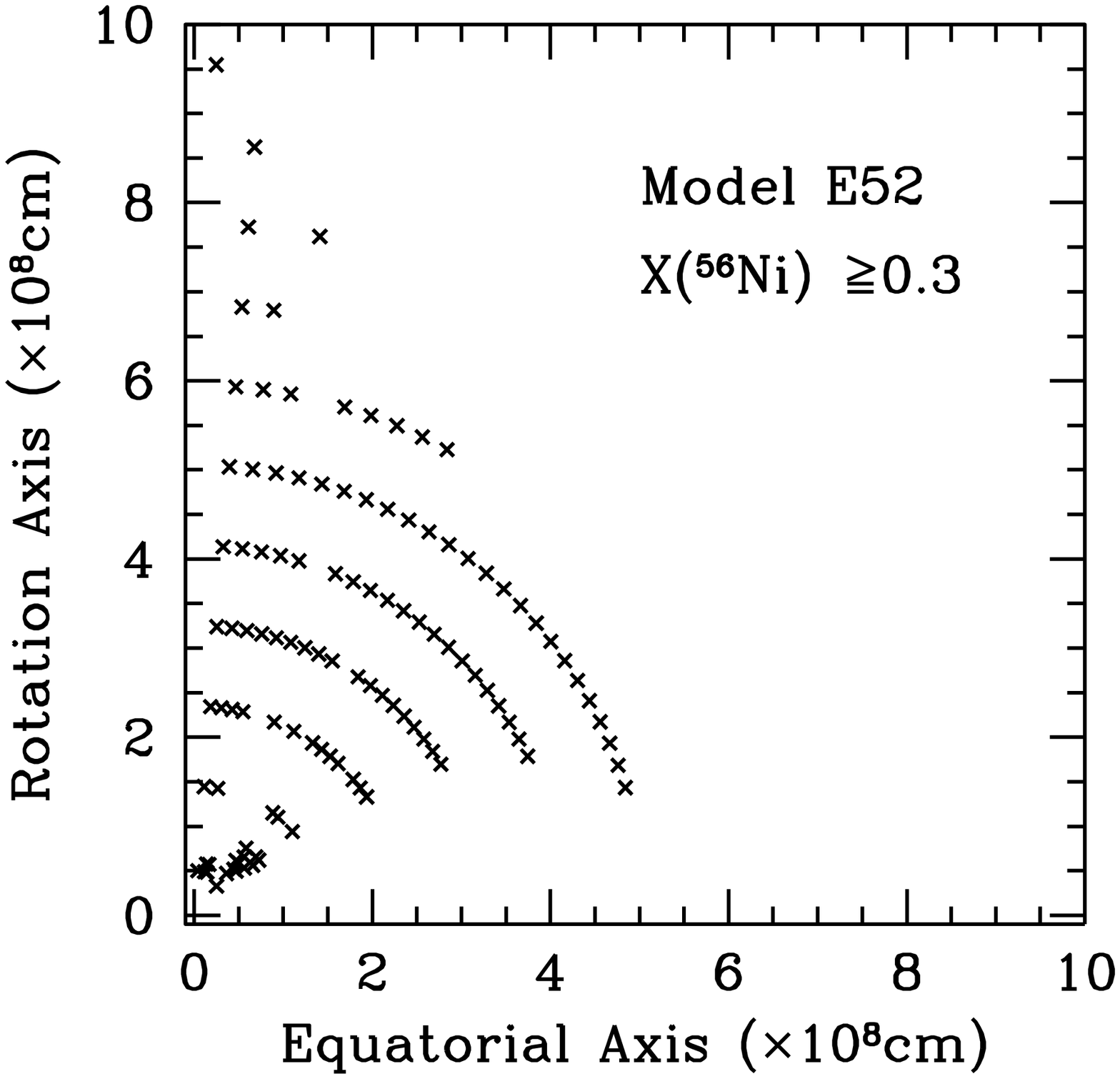}{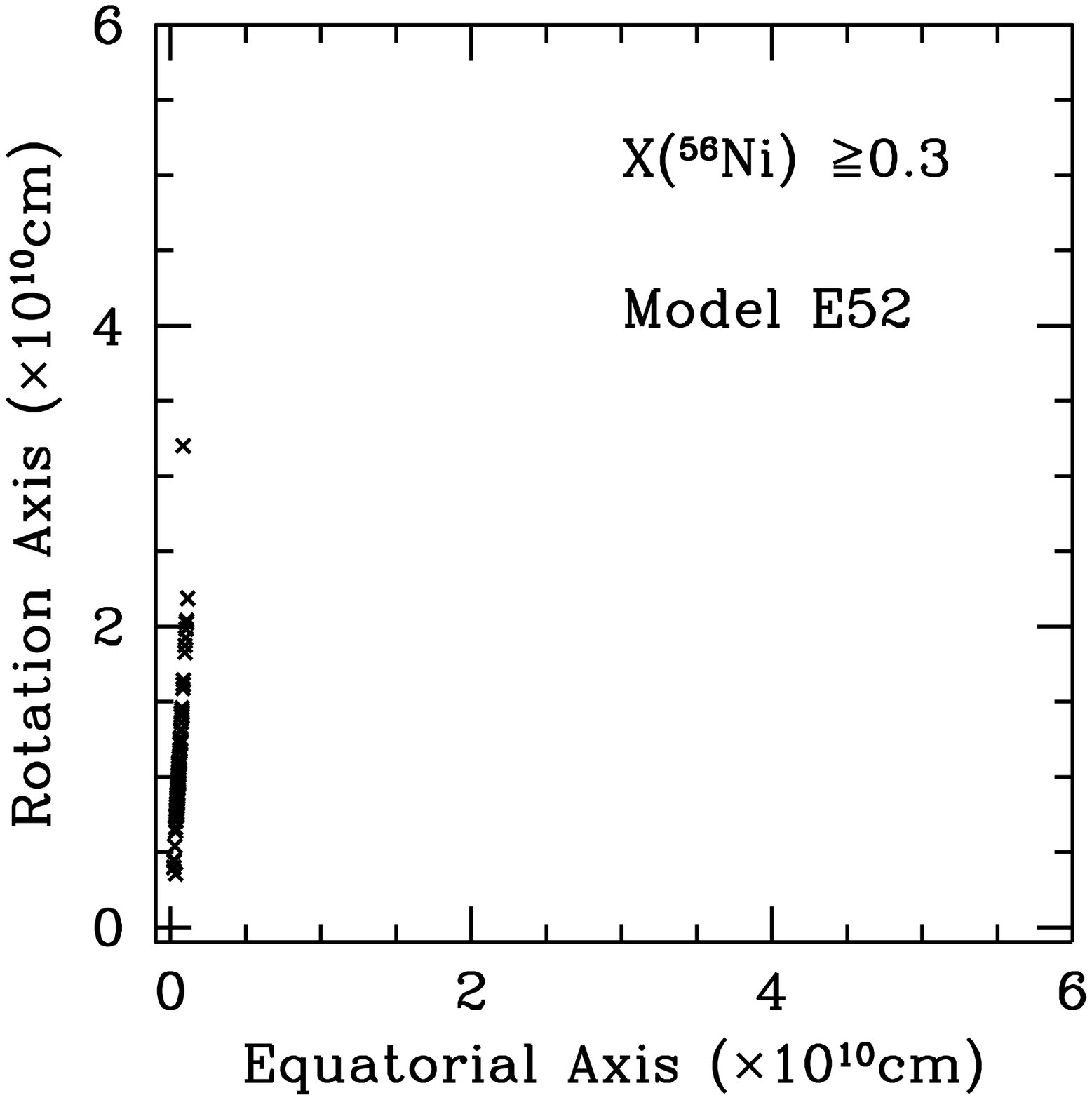}
\caption{Same with Fig.~\ref{fig8}, but for Model E52.
Total ejected mass of $\rm ^{56}Ni$
is 0.23$M_{\odot}$ (Model E52), which is comparable to the
observed value of hypernovae. However, most of the synthesized
$\rm ^{56}Ni$ is in the jet component (0.23$M_{\odot}$), while
small amount of $\rm ^{56}Ni$ (0.00229$M_{\odot}$) is in the
supernova component.
 \label{fig9}}
\end{figure}

\begin{figure}
\epsscale{1.0}
\plottwo{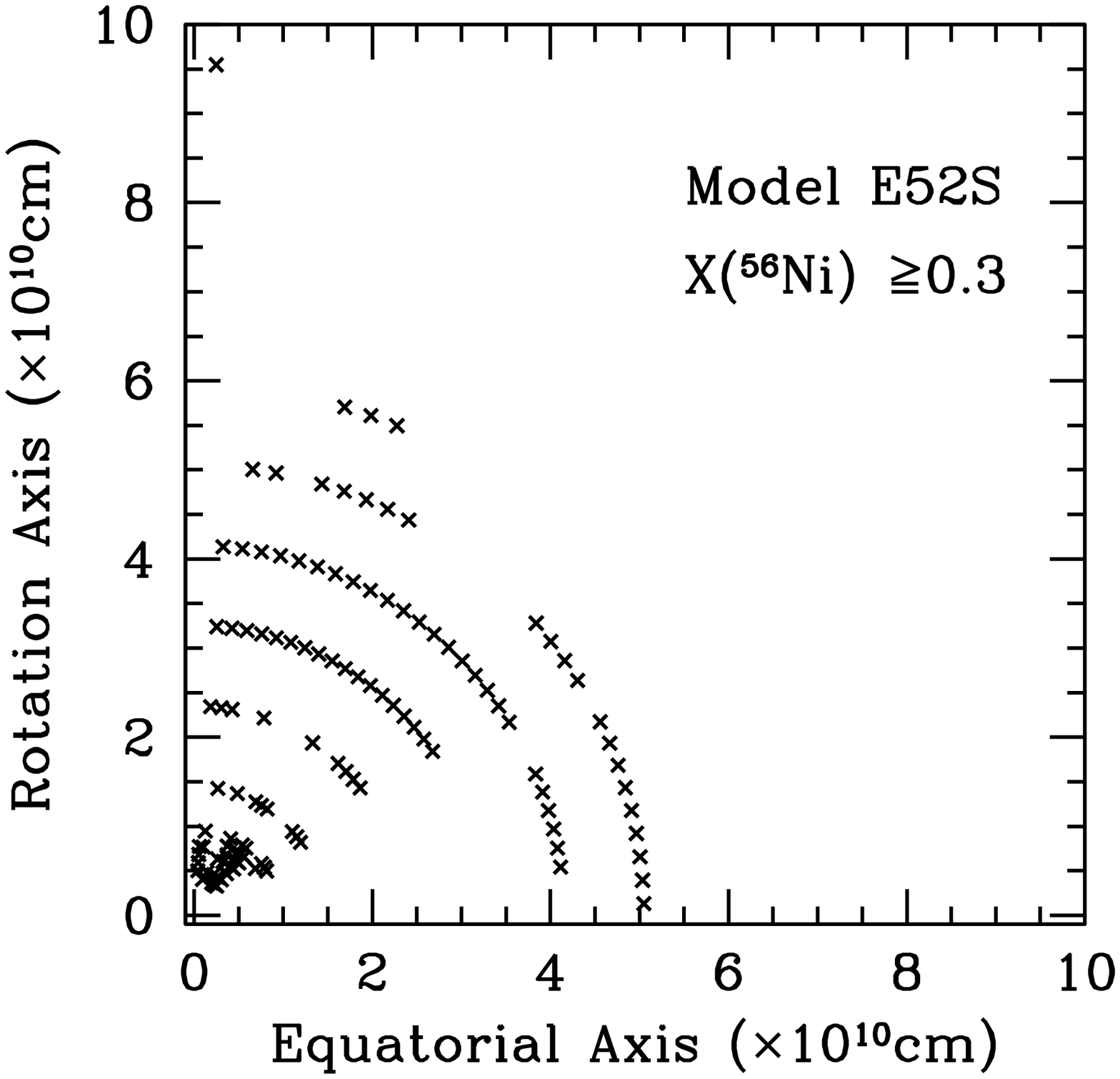}{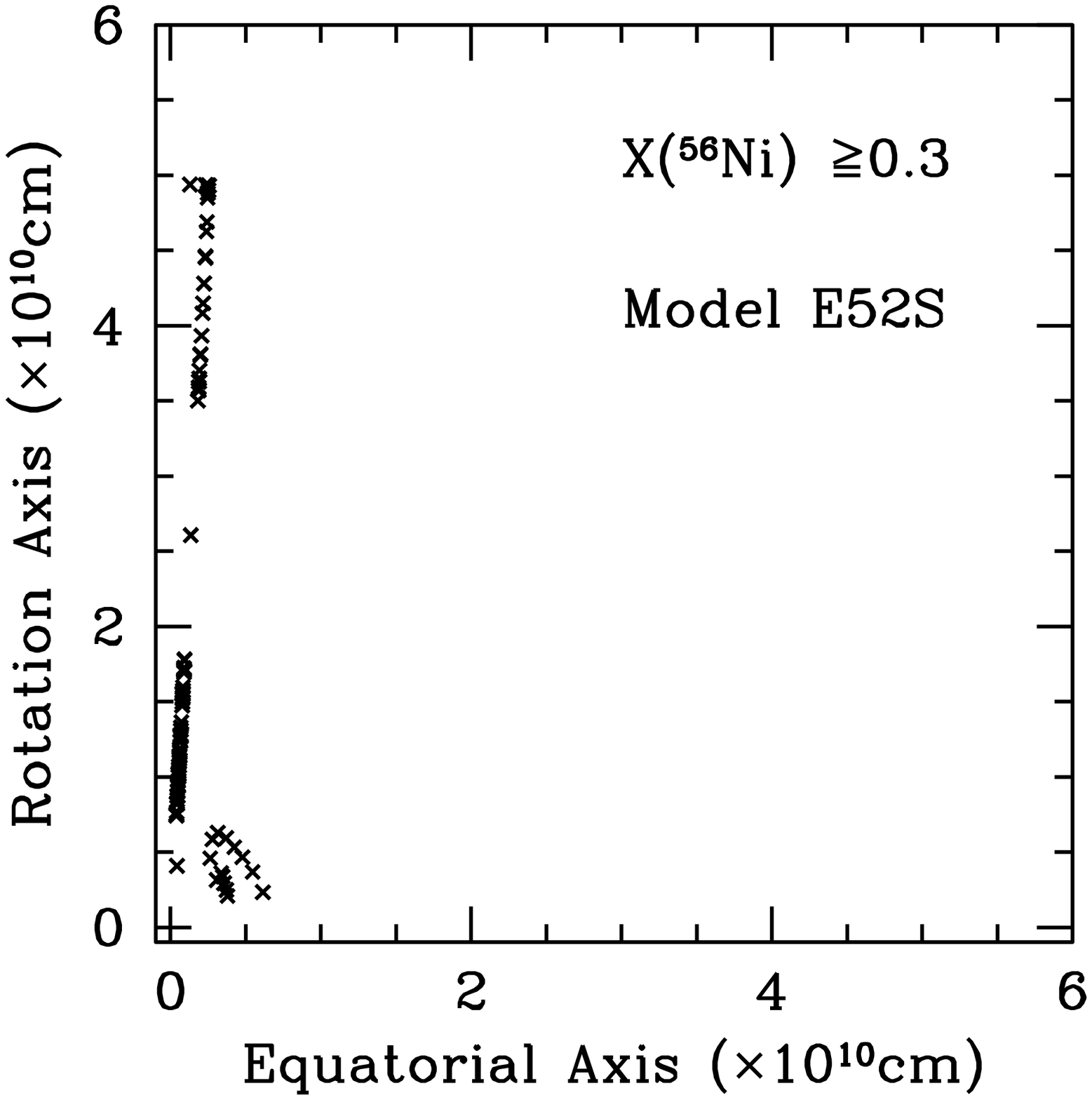}
\caption{Same with Fig.~\ref{fig8}, but for Model E52s (right panel).
Total ejected mass of $\rm ^{56}Ni$ is 0.28$M_{\odot}$ (Model E52s),
respectively, which is comparable to the observed value of hypernovae.
However, as Model E52, most of the synthesized
$\rm ^{56}Ni$ is in the jet component (0.195$M_{\odot}$), while
small amount of $\rm ^{56}Ni$ (0.0882$M_{\odot}$) is in the
supernova component.
 \label{fig10}}
\end{figure}

We can know where and how much $\rm ^{56}Ni$ is synthesized by doing
post processing. Also, we can see how $\rm ^{56}Ni$ is ejected, that
is, how much $\rm ^{56}Ni$ is ejected as jet component or SN component.
In Fig.~\ref{fig8}, positions of the ejected test particles for Model
E51 at $t=0$ sec (left panel) and $t= 4.27$ sec (right panel) are shown
that satisfy the condition that the mass fraction of $\rm ^{56}Ni$ becomes
greater than 0.3 as a result of explosive nucleosynthesis.
Total ejected mass of $\rm ^{56}Ni$
becomes 0.0439$M_{\odot}$. In particular, total mass of it in the
SN component is 0.0175$M_{\odot}$, which is much smaller than 
the observed values of hypernovae. In Figs.~\ref{fig9} and
\ref{fig10}, same figures with Fig.~\ref{fig8} but for Model E52
and E52S are shown. 
As mentioned above, the outflow becomes bi-polar due to the
asymmetry of density structure. As a result, jet component
can be seen even in Model E52S (Fig.~\ref{fig10}).
Total ejected mass of $\rm ^{56}Ni$
is 0.23$M_{\odot}$ (Model E52), which is comparable to the
observed values of hypernovae. 
It is also noted that most of the synthesized
$\rm ^{56}Ni$ is in the jet component (0.23$M_{\odot}$), while
small amount of $\rm ^{56}Ni$ (0.00229$M_{\odot}$) is in the
SN component. From this result, we can guess that some fraction
of gamma-ray lines from $\rm ^{56}Ni$ decays appear
without losing their energies. This point is discussed in
section~\ref{discussion} in detail.
Of course, it is noted this model will not explain association of
GRBs with hypernovae since this model can not cause highly
relativistic jets. 
The features of Model E52S are same with E52. 
That is, the ejected mass of $\rm ^{56}Ni$ in Model E52S is
0.28$M_{\odot}$, which is comparable to Model E52 and is much larger
than Model E51. From this result, it is shown that the total ejected 
mass of $\rm ^{56}Ni$ depends sensitively on the energy deposition
rate, and depends not so sensitively on the mass of the heated region
(energy-injection region), which is consistent with Nagataki et
al. (2003). The reason is as follows:
The criterion for the complete silicon burning is $T_{\rm max} \ge
5\times 10^9$ [K]~\cite{thielemann96}. 
It is well known that the matter behind the shock wave is radiation
dominated and $T_{\rm max}$ can be estimated well by equating the
supernova (hypernova) energy with the radiation energy inside the
radius $r$ of the shock front
\begin{eqnarray}
E_{\rm HN} = 10^{52} \left(\frac{E_{\rm HN}}{10^{52} \rm erg }
\right)
           = \frac{11 \pi^3}{45} \frac{k_b^4}{\hbar^3 c^3} r^3 T_{\rm
           max}^4 \;\;\; \rm [erg],
\end{eqnarray}
where $E_{\rm HN}$ is the total explosion energy of a hypernova,
$\hbar$ is the Planck constant divided
by $2 \pi$, $c$ is speed of light. Here spherical
explosion is assumed. This equation gives $r$ as
\begin{eqnarray}
r = 5.7 \times 10^8 \left( \frac{5 \times 10^9 \rm K}{T_{\rm max}}
\right)^{4/3} 
\left( \frac{E_{\rm HN}}{10^{52} \rm erg} \right)^{1/3} \;\;\; \rm [cm].
\end{eqnarray}
In the case of Model E52S, $\rm ^{56}Ni$ is synthesized within the edge
for the complete silicon burning ($\sim 6 \times 10^8$ cm, see left
panel of Fig.~\ref{fig10}). The situation should be almost same in
the jet-induced explosion Models (Model E52, see left panel of
Fig.~\ref{fig9}). On the other hand, in the case of Model E51, matter
starts to move outwards after the passage of the shock wave, and
almost all of the matter move away ($r \ge 6\times 10^8$ cm) before
the injection of thermal energy (= $10^{52}$ erg) is completed. This
is the reason why abundance of $\rm ^{56}Ni$ is little in Model E51. 
As mentioned above, $\rm ^{56}Ni$
is synthesized within $r \le 10^9$ cm. So we can conclude that
the amount of $\rm ^{56}Ni$ does not depend on the assumption that
the region ($\ge 10^{10}$ cm) where the shock wave does not reach
even at the final stage of calculation ($t=10$ sec) can escape to
infinity. It should be noted
that the chemical composition in the jet is not unchanged for the
matter located at $r \ge 10^9$ cm initially. So helium layer and outer oxygen
layer located at $r \ge 10^9$ cm are ejected as a jet with chemical
composition unchanged.

\placefigure{fig8}
\placefigure{fig9}
\placefigure{fig10}

\begin{figure}
\epsscale{1.0}
\plottwo{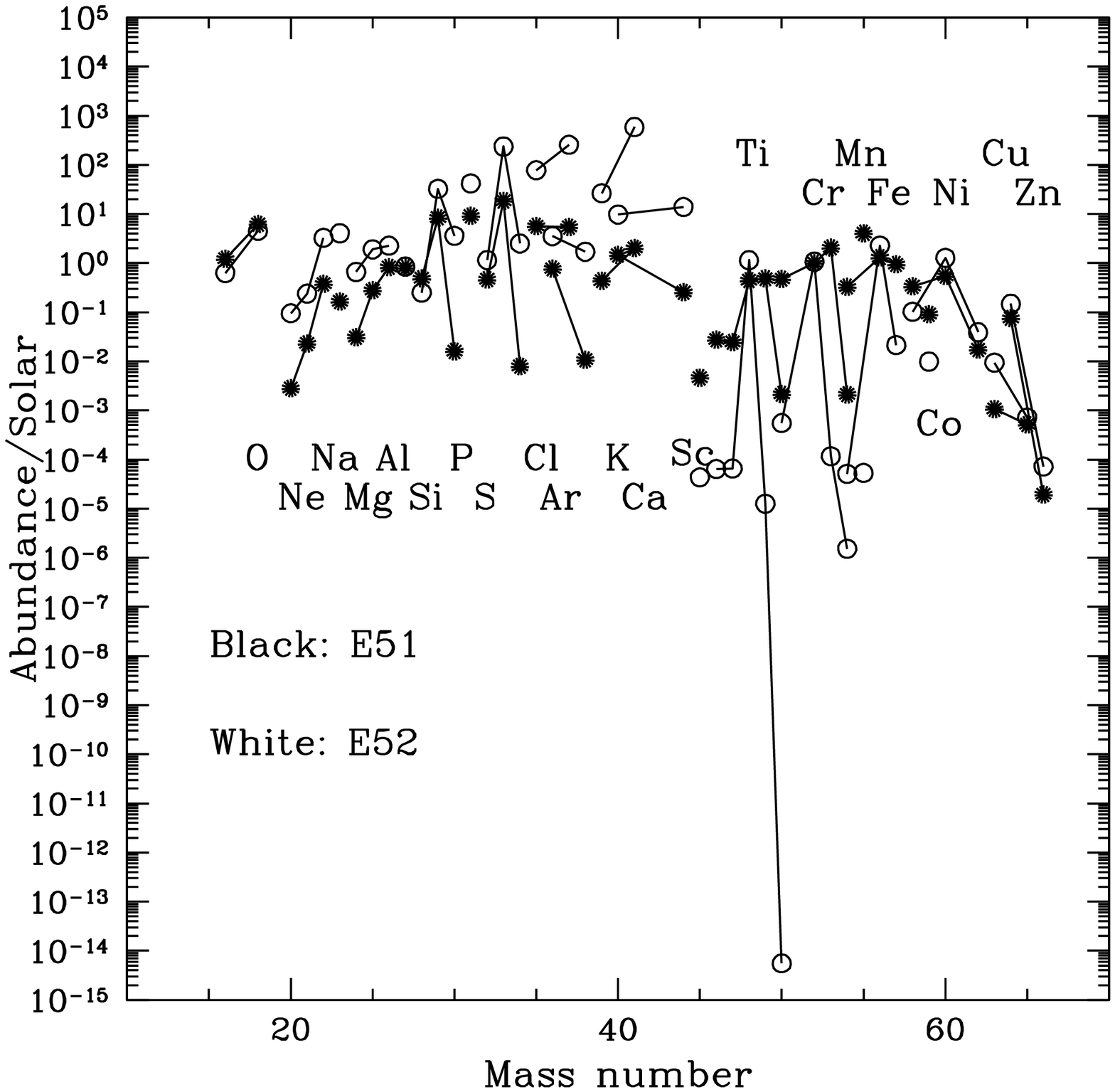}{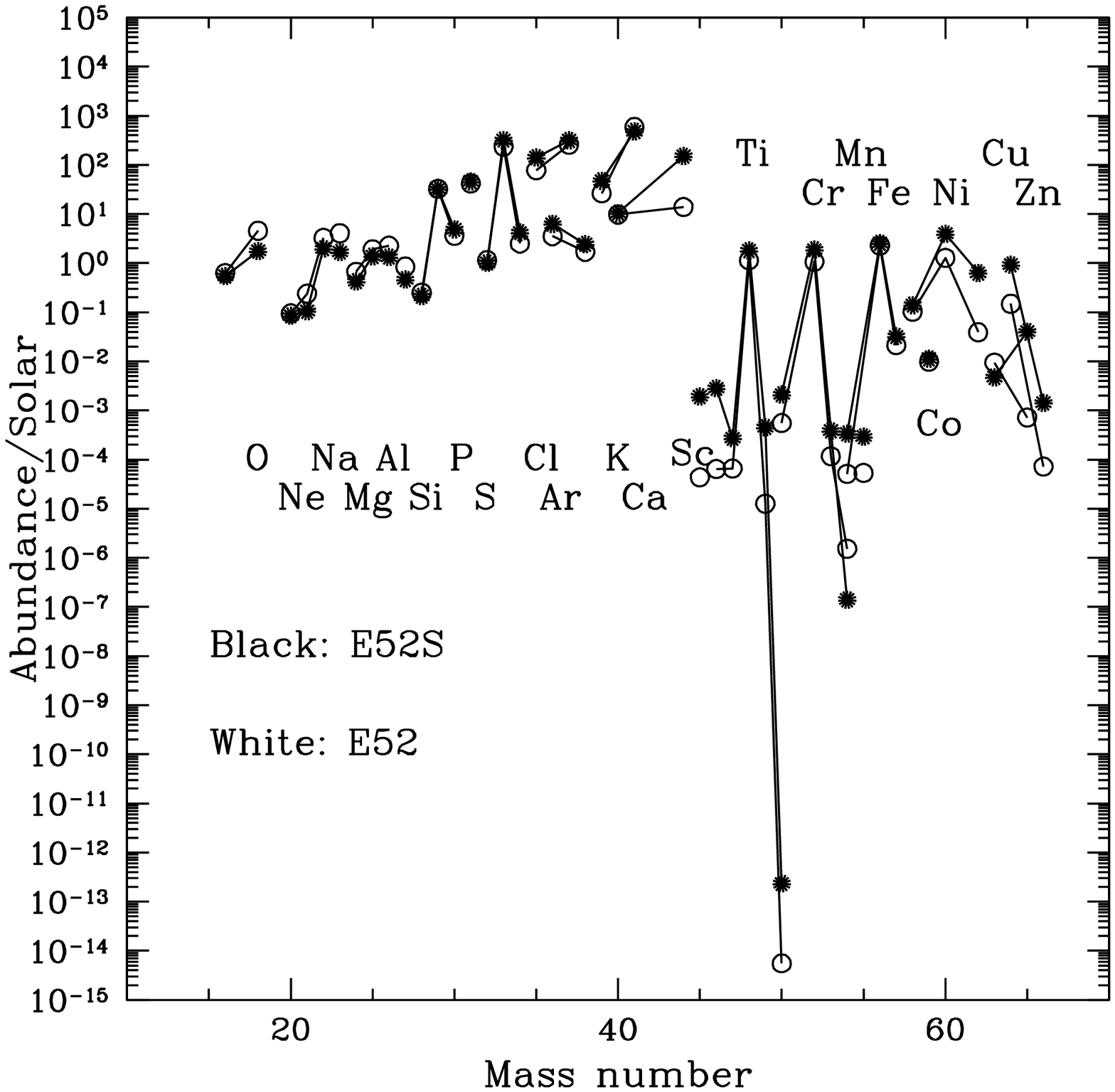}
\caption{Abundance of heavy elements in the ejecta normalized by the solar
value. All unstable nuclei produced in the ejecta are assumed to decay
 to the corresponding stable nuclei. Models E51 (black) and E52 (white) are shown in
 the left panel while Models E52S (black) and E52 (white) are shown in the right panel.
 \label{fig11}}
\end{figure}

We have calculated the abundance of heavy elements in the ejecta,
using the mass cut and post processing mentioned above. The result is
shown in Table~\ref{tab2}.
Abundances are written in units of $M_{\odot}$. All unstable nuclei produced
in the ejecta are assumed to decay to the corresponding stable
nuclei. The amount of $\rm ^{56}Ni$ is also shown in the last row.
As explained above, the amount of $\rm ^{56}Ni$ does not depend on
the assumption that the region where the shock wave does not reach
even at the final stage can escape to infinity. However,
the amounts of light elements such as $\rm ^{16}O$ and $\rm ^{4}He$
depends on this assumption, because the progenitor is composed of
such light elements. So we think further study should be required
to estimate the abundance of light elements ejected from a collapsar
more precisely. The abundance of heavy elements in the ejecta
normalized by the solar value~\citep{anders89,nagataki99}
is shown in Fig.~\ref{fig11}.
Models E51 (black) and E52 (white) are shown in the left panel,
while Models E52S (black) and E52 (white) are shown in the right panel.
We can see that there is an enhancement of nuclei whose mass number $\sim
40$ in Models E52 and E52S. This is the result of  incomplete
silicon burning and alpha-rich
freezeout~\citep{nagataki97,nagataki98b,nagataki00,maeda02,maeda03}.

\placefigure{fig11}
\placetable{tab2}

\section{DISCUSSION}\label{discussion}

First, we discuss the formation of highly relativistic jets to
realize GRBs. Of course, in the present study, we can not investigate
the acceleration of the jet to relativistic regime since our numerical
code is Newtonian. So we have to investigate this topic using
relativistic hydrodynamic code. At present, we discuss this topic by
introducing previous works that investigate GRB jets using
relativistic hydrodynamic code. Aloy et al. (2000) performed such a
calculation using the collapsar model~\cite{macfadyen99}. They have
shown that the jet is formed as a consequence of an assumed energy
deposition rate in the range of $10^{50} - 10^{51}$ erg s$^{-1}$ within a
30$^{\circ}$ cone around the rotation axis, which is similar treatment
of Model E51 in this study. They reported that the maximum Lorentz factor of
the jet becomes 44, which seems to be smaller than required value
to explain GRBs ($\sim$ 300; Piran 1999). Zhang et al. (2003) also
calculated propagation of the relativistic jet through the collapsar
with constant energy deposition rate $10^{50} - 3 \times 10^{50}$
erg s$^{-1}$.
They set the location of the inner boundary to be $2 \times 10^8$ cm. 
They estimated terminal Lorentz factor, which is
calculated by assuming that all internal energy is converted into
kinetic energy, becomes $\sim$100, although their calculated Lorentz
factor of the jet is $\sim$50 at most. Zhang and Woosley (2004) have
improved their code and demonstrated that the bulk Lorentz factor of
the jet can reach $\sim$100, although they set the inner boundary to
be $10^{10}$ cm. From their works, we can understand it very difficult
to realize a highly relativistic jet whose bulk Lorentz factor becomes
larger than 100. They required a collimated, narrow jet by depositing
explosion energy for $\sim 10$ sec. The importance of the collimation
to realize a highly relativistic jet can also be understood by 
rough estimation.
We can calculate the mass of the progenitor included
within a cone around the rotation axis. The masses included within a
cone with the zenith angles $3^{\circ},5^{\circ},10^{\circ},15^{\circ}$ 
are $7.9 \times 10^{30}, 3.1 \times 10^{31}, 7.1 \times 10^{31}, 2.0
\times 10^{32}$ g, respectively. So if these matter are accelerated to
highly relativistic regime with bulk Lorentz factor $\Gamma$, the
kinetic energies have to be $7.1 \times 10^{53}(\Gamma/100) , 2.8
\times 10^{54}(\Gamma/100), 6.4 \times 10^{54}(\Gamma/100), 1.8 
\times 10^{55}(\Gamma/100)$ erg, respectively, which shows the importance of
collimation.
The importance of collimation is also confirmed by our forth-coming
paper~\cite{mizuta06}.   
In Models E52 and E52S, the opening angles of the
jets are wider than that in Model E51 (see
Figs.~\ref{fig2},\ref{fig3},\ref{fig4}),
so we consider that highly relativistic jet will not be produced in
Models E52 and E52S even if relativistic hydrodynamic code is used.
So we consider that Models E52 and E52S can not explain the phenomena
of association of GRBs with hypernovae, even though much $\rm ^{56}Ni$
is synthesized in these models.

There is another question. 
Does all of $\rm ^{56}Ni$ produced in the jet
of the collapsar model brighten the supernova remnant?
If the jet becomes optically thin before $\rm ^{56}Ni$ decays into
$\rm ^{56}Co$ and $\rm ^{56}Co$ decays into $\rm ^{56} Fe$, these
nuclei should result in only emitting gamma-rays and can not
brighten the supernova remnant. Colgate et al. (1980) consider the
deposition of energy by gamma-rays emanating from decay of $\rm ^{56}Ni$
and $\rm ^{56}Co$. They found that the mass opacity of the gamma-ray
absorption is about 0.029 cm$^2$ g$^{-1}$, for either $\rm ^{56}Ni$
or $\rm ^{56}Co$ decay spectrum. The half-lives of $\rm ^{56}Ni$
and $\rm ^{56}Co$ are 5.9 and 77.1 days. So we can roughly estimate the
opacity for these gamma-rays in the jet.
We assume that the opening angle of the jet is 10$^{\circ}$ and
expansion velocity of the jet is speed of light. Under these assumption,
when substantial $\rm ^{56}Ni$ decays into $\rm ^{56}Co$, the volume of the
jet becomes $3.2 \times 10^{-2} R^3$ cm$^3$, where $R$ is the radius
of the jet with $R=1.5 \times
10^{16} \Gamma $ cm and $\Gamma$ is the bulk Lorentz factor of the jet. 
Since the mass included within a cone with the zenith angles 
$10^{\circ}$ is $7.1 \times 10^{31}$ g, the density of the jet at that
time becomes $6.6 \times 10^{-16}$ g cm$^{-3} \Gamma^{-3}$.
So the mean free path of the gamma-rays becomes $5.3 \times 10^{16}
\Gamma^3$ cm, which is comparable or longer than the radius of the
jet. Of course, the gamma-rays can easily escape from the side of the
cone. So we think most of the gamma-rays produced in a jet will escape without
depositing energy to the jet and supernova components as long as the
bulk speed of the jet is relativistic. Similar discussion 
can be adopted for the decays of $\rm ^{56}Co$ into $\rm ^{56}Fe$.
So, we can conclude that some fraction of gamma-ray lines from
$\rm ^{56}Ni$ decays in the jet
may 
appear as gamma-rays, which may
be observed as relativistically Lorentz boosted gamma-ray line
profiles in future. 
To obtain more firm conclusion, it will be necessary to perform
multi-dimensional relativistic hydrodynamics with radiation transfer
and calculate the light curve of hypernovae.

Here we have to make a comment on the resolution of numerical
simulations in this study. It is shown in numerical simulations
with high resolution~\cite{zhang04} that the jet propagates
only mildly relativistically ($\sim$c/2) while in the star, and
shocked gas can move laterally to form a cocoon, allowing the core
of the jet to remain relativistic. 
Thus the total mass accelerated
relativistically ($\Gamma \sim 100$) by the jet is not the fraction
of the star intercepted by the jet, although the importance of
collimation can be understood by calculating mass within the cone
of the jet as mentioned above. So, strictly speaking, there
are three components in a collapsar model, highly relativistic
jet, cocoon, and supernova component when relativistic hydrodynamic
simulations are performed. In this study, highly
relativistic jet and cocoon are called as 'jet component'.

We concluded that the highly relativistic
jet and mildly relativistic cocoon are optically thin against the gamma-rays that come
from decays of $\rm ^{56}Ni$ and $\rm ^{56}Co$. 
As mentioned above, mean free path of the gamma-rays becomes 
$5.3 \times 10^{16} \Gamma^3$ cm, which is comparable or longer than
the radius of the jet component, $1.5 \times 10^{16} \Gamma $ cm
(Note that these are comparable even if $\Gamma = 1$).
So the $\rm ^{56}Ni$ in the highly relativistic jet and mildly
relativistic cocoon
should not contribute to the light curve of a supernova.
On the other hand, sub-relativistic cocoon will be optically thick
against the gamma-rays. So sub-relativistic cocoon may have
contribution to the optical light curve of hypernovae.
In fact, it is clear that too much energy is required to
accelerate all of $\rm ^{56}Ni$ in the jet to relativistic
speed. From the discussion of energetics mentioned above,
the required energy (erg) becomes $9.0 \times 10^{53} (\Gamma/1)
(M_{\rm Ni}/0.5M_{\odot})$ where $M_{\rm Ni}$ is the mass of $\rm ^{56}Ni$. 
Thus we consider that there is a possibility that a part of the
gamma-ray lines may appear without losing their energies.
As mentioned above, numerical simulations of relativistic
hydrodynamics with high resolution should be required to distinguish
highly relativistic jet from cocoon, and to obtain precise 
distribution of burning products, which we are planning to
simulate in near future.

Let us emphasize our motivation in this work here.
In this study, we want to consider the consistency between the 
collapsar model of MacFadyen and Woosley (1999) and explosive
nucleosynthesis in a hypernova jet. The answer is as follows.
From the discussions mentioned above, it seems difficult to explain
the required amount of $\rm ^{56}Ni$ ($\sim 0.5 M_{\odot}$) by the
explosive nucleosynthesis in the jet. We think the another idea that
the origin of $\rm ^{56}Ni$ is the explosive nucleosynthesis in the
accretion disk~\citep{macfadyen99,pruet02} is much simpler and
adequate to explain the association of GRBs and hypernovae. 
In this scenario, it is not necessary for $\rm ^{56}Ni$ to be
synthesized in a short timescale as Models E52 and E52S. As explained
in section~\ref{intro}, in the collapsar scenario, the jet is launched
$\sim 7$ sec after the gravitational collapse and the duration of the
jet is about 10 sec, which is much longer than the typical timescale
of normal core-collapse supernovae and comparable to the typical
observed duration of GRBs.
As a result of gravitational collapse in a long timescale,
the density around the rotation axis becomes low, which is a good
environment to produce a fire ball. That is, long
timescale of the order of 10 sec is essential to realize a GRB. As shown
in this study, in such a case (Model E51), $\rm ^{56}Ni$ is not
synthesized so much in the jet. Rather, it will be natural to
consider that the origin of
$\rm ^{56}Ni$ is the accretion disk around the black hole. In this
scenario, $\rm ^{56}Ni$ is also ejected in a long timescale of the order
of 10 sec. No requirement that $\rm ^{56}Ni$ has to be produced in a
short timescale exists in this scenario.
Also, as shown in this study, the explosion becomes
naturally bi-polar in any case (even in Model E52S) due to the
aspherical density structure of the progenitor. So it will be natural
to consider that the $\rm ^{56}Ni$ synthesized
in the accretion disk and conveyed as outflows are blown along to
the rotation axis, which can explain the line features of SN 1998bw
and double peaked line features of SN 2003jd~\cite{mazzali05}. 
Of course, there is much uncertainty how much $\rm ^{56}Ni$
is ejected from the accretion disk. This problem depends sensitively
on the effects of viscosity. Further investigation is required to
estimate how much $\rm ^{56}Ni$ is ejected.
At present study too, how much $\rm ^{56}Ni$ is ejected from the accretion
disk is not estimated since artifitial viscosity and/or magnetic
fields are not included.

It should be noted that there are varieties of observations and theories
related with the association of GRBs with supernovae. As a result, it 
will be natural to consider that there are varieties of explosive
nucleosynthesis in the collapsar. For example, there will be a class
of 'failed GRBs'~\citep{lazzati02,totani03}, in which baryon-rich jet
propagates. In this class, there is a possibility that much heavy
elements are synthesized because of high density in the
jet~\cite{inoue04}, while light elements will be synthesized in
baryon-poor jets~\citep{lemoine02,pruet02,beloborodov03}.
Also, if the central engine of the GRBs is a
magnetar~\citep{rees00,takiwaki04} or
magnetized
collapsar~\citep{blandford77,blandford82,proga03,koide03,mizuno04,proga05,mckinney05a,mckinney05b},
the timescale of the explosion will be
shorter than that of a collapsar, which will result in the different
way of nucleosynthesis in this study. Further investigation should
still be required to understand the central engine of GRBs and origin of
$\rm ^{56}Ni$ in hypernovae.

In the early universe, where the metal content of gas is very low, the
enrichment by a single supernova can dominate the preexisting metal
contents~\cite{audouze95}. Since GRBs also occurs from the early
universe, there is a possibility that some fraction of metal poor star
reflects the chemical abundance of GRBs accompanied by
hypernovae~\cite{maeda03}. From Fig.~\ref{fig11}, we can see that there
is an enhancement of nuclei whose mass number $\sim 40$ in Models E52 and
E52S compared with Model E51. This is the result of  incomplete
silicon burning and alpha-rich
freezeout~\citep{nagataki97,nagataki98b,nagataki00,maeda02,maeda03}.
In particular, in Models E52 and E52S, [Ca/Si] $\equiv \log
(X_{\rm Ca}/X_{\rm Si}) - \log (X_{\rm Ca}/X_{\rm Si})_{\odot}$ is larger
than unity, where $X_{i}$ is the mass fraction of the $i$th element
and $(X_{\rm Ca}/X_{\rm Si})_{\odot}$ is the solar value, which is in
contrast with Model E51 and previous works~\citep{qian02,pruet04}.
So it may be determined which model is the proper one as a model
of hypernova
by observations of chemical composition in metal poor
stars~\citep{ishimaru99,umeda02,tsujimoto04,ishimaru04,umeda05}.

It will be a good challenge to perform a calculation of the
r-process and/or p-process nucleosynthesis in the GRB jet in this
study, because this jet will also be able to realize a high entropy
condition enough to realize these
processes~\citep{macfadyen99,nagataki00,nagataki01,nagataki01b,wanajo02,suzuki05}.          
As shown in Fig.~\ref{fig7}, there is really a region in the jet where
high entropy per baryon is realized (it reaches to $10^5$ at most!).
However, in this study, test particles in the downstream of the shock
wave are moving with shock velocity, and no test particles are left inside of
the shocked region where entropy per baryon is quite high. So other
method will be required to investigate the r-process and/or p-process
nucleosynthesis in the jet.
We are planning to perform such calculations. Results will be
presented in near future.

After we have written this manuscript, we found recent papers
by Maeda et al. (2005) and Maeda (2005) in which similar topic of this
study is considered. Their conclusions are
consistent with our prediction. That is, they calculated
gamma-ray lines from $\rm ^{56}Ni$ decays in the jet, although relativistically
Lorentz boosted line profiles are not calculated because they used a
Newtonian code. Also, it is noted that they deposit all explosion
energy as an initial condition like Model E52 and E52S in this study.
That is why they can obtain enough $\rm ^{56}Ni$ to explain the light
curve of SN 1998bw, which is consistent with the results in this
study. The reason why they obtained a little more $\rm ^{56}Ni$ ($\sim
0.4M_{\odot}$) may be the difference of the progenitor. 
They used a spherically symmetric progenitor while we have
used a collapsar model. As stated in this study, the density of the
collapsar model around the polar region is lower than the spherically symmetric
progenitor. So the abundance of synthesized $\rm ^{56}Ni$ may be
smaller in this study. 

\section{SUMMARY AND CONCLUSION}\label{conclusion}

We have performed 2-dimensional hydrodynamic simulations to investigate
explosive nucleosynthesis in a collapsar using the model of MacFadyen
and Woosley (1999).
We have shown that $\rm ^{56}Ni$ is not produced in the jet
sufficiently to explain the observed amount of a hypernova such as
SN 1998bw when the duration
of the explosion is $\sim$10 sec (the standard model, E51). 
Even though a considerable amount of $\rm ^{56}Ni$ is
synthesized if all explosion energy is deposited initially (the
extreme models, E52 and E52S), the opening angles of the jets become
too wide to realize highly relativistic outflows and a GRB in such a case.
From these results, it is concluded that the origin 
of $\rm ^{56}Ni$ in hypernovae associated with GRBs is not the explosive
nucleosynthesis in the jet. We consider that the idea that
the origin of $\rm ^{56}Ni$ in hypernovae is the explosive
nucleosynthesis in the accretion disk is more promising.
We have also shown that the explosion
becomes bi-polar naturally due to the effect of the deformed
progenitor. This fact suggests that the $\rm ^{56}Ni$ synthesized
in the accretion disk and conveyed as outflows are blown along to
the rotation axis, which will explain the line features of SN 1998bw
and double peaked line features of SN 2003jd, and will help the idea
of the accretion disk mentioned above.

Also, some predictions are presented in this study.
Some fraction of the gamma-ray lines from $\rm ^{56}Ni$ decays in the
jet will appear without losing their energies because the jet
becomes optically thin before a considerable amount of $\rm ^{56}Ni$
decays as long as the jet is a relativistic flow. So
it has been predicted that some
fraction of $\rm ^{56}Ni$ synthesized in the jet may show
relativistically Lorentz
boosted line profiles. That is, highly blue shifted (or red shifted)
broad line features might be observed in future. 
It has been also shown that there
is an enhancement of nuclei whose mass number $\sim 40$ in Models E52 and
E52S compared with Model E51 as a result of incomplete
silicon burning and alpha-rich freezeout.
So it may be determined which model is the proper one as a
model of a gamma-ray burst accompanied by a hypernova by observations
of chemical composition in metal poor stars.

Of course there is still uncertainty of observations and theories
on the formation of GRBs accompanied by hypernovae.
Further investigation should still be required to understand
the central engine of GRBs and origin of $\rm ^{56}Ni$ in hypernovae.

\acknowledgments
We are grateful to the anonymous referee of this paper for
fruitful comments.
We are grateful to A. MacFadyen for giving us a result of numerical
simulation of a collapsar and useful discussion.
S.N. are also grateful to M. Watanabe and S. Yamada for useful discussion.
The computation was partly carried out on NEC SX-5 and SX-8, SGI
Altix3700 BX2, and Compaq
AlphaServer ES40 at Yukawa Institute for Theoretical Physics, and
Fujitsu VPP5000 at National Astronomical Observatory of Japan.
This work is in part supported by a Grant-in-Aid for the 21st Century
COE ``Center for Diversity and Universality in Physics'' from the
Ministry of Education, Culture, Sports, Science and Technology of
Japan. S.N. is partially supported by Grants-in-Aid for Scientific
Research from the Ministry of Education, Culture, Sports, Science and
Technology of Japan through No. 14102004, 14079202, and 16740134.

\clearpage

\begin{table*}
\begin{center}
\begin{tabular}{ccccccccccccccccc}
\tableline
\tableline
Model & $\theta_{\rm Jet}$ & $\dot{E}$ [erg s$^{-1}$] & $E_{\rm tot}$ [erg] \\
\tableline
E51 & 30$^{\circ}$ & 10$^{51}$ & $10^{52}$  \\
E52 & 30$^{\circ}$ & $\infty$    & $10^{52}$  \\
E52S& 90$^{\circ}$ & $\infty$    & $10^{52}$  \\
\tableline
\end{tabular}
\tablenum{1}
\caption{
Models, half-angle of the initial jet, thermal energy deposition rate,
and total explosion energy.  
}\label{tab1}
\end{center}
\end{table*}

\begin{table*}
\begin{center}
\begin{tabular}{ccccccccccccccccc}
\tableline
\tableline
Element & $\rm A_{min}$ & $\rm A_{max}$  & Element & $\rm A_{min}$ &
$\rm A_{max}$ & Element & $\rm A_{min}$ & $\rm A_{max}$ \\
\tableline
N & 1 & 1 & Al& 24 & 30 & V & 44 & 54 \\
H & 1 & 1 & Si& 26 & 33 & Cr& 46 & 55 \\
He& 4 & 4 & P & 28 & 36 & Mn& 48 & 58 \\
C & 11& 14& S & 31 & 37 & Fe& 52 & 61 \\
N & 12& 15& Cl& 32 & 40 & Co& 54 & 64 \\
O & 14& 19& Ar& 35 & 45 & Ni& 56 & 65 \\
F & 17& 22& K & 36 & 48 & Cu& 58 & 68 \\
Ne& 18& 23& Ca& 39 & 49 & Zn& 60 & 71 \\
Na& 20& 26& Sc& 40 & 51 & Ga& 62 & 73 \\
Mg& 22& 27& Ti& 42 & 52 & Ge& 64 & 74 \\
\tableline
\end{tabular}
\tablenum{2}
\caption{
Nuclear Reaction Network Employed
}\label{tabnucl}
\end{center}
\end{table*}

\clearpage
\begin{table*}
\small
\begin{center}
\begin{tabular*}{140mm}{ccccccccccccccccccc}
\hline
\hline
Species & E51 & E52 & E52S & Species & E51 & E52 & E52S \\
\hline
$\rm ^{16}O$  &3.38E-01 &5.27E-01  &5.01E-01 &$\rm ^{45}Sc$ &7.33E-09 &3.61E-10  &1.96E-08 \\
$\rm ^{18}O$  &3.93E-03 &8.62E-03  &3.66E-03 &$\rm ^{46}Ti$ &1.79E-07 &1.25E-09  &6.01E-08 \\
$\rm ^{20}Ne$ &1.33E-04 &1.34E-02  &1.30E-02 &$\rm ^{47}Ti$ &1.48E-07 &1.19E-09  &5.39E-09 \\
$\rm ^{21}Ne$ &2.74E-06 &8.55E-05  &4.15E-05 &$\rm ^{48}Ti$ &2.77E-05 &2.15E-04  &3.68E-04 \\
$\rm ^{22}Ne$ &1.43E-03 &3.70E-02  &2.45E-02 &$\rm ^{49}Ti$ &2.30E-06 &1.79E-10  &7.01E-09 \\
$\rm ^{23}Na$ &1.60E-04 &1.18E-02  &5.33E-03 &$\rm ^{50}Ti$ &1.02E-08 &7.94E-20  &3.67E-18 \\
$\rm ^{24}Mg$ &4.63E-04 &2.97E-02  &2.08E-02 &$\rm ^{50}Cr$ &1.01E-05 &3.58E-08  &1.49E-07 \\
$\rm ^{25}Mg$ &5.51E-04 &1.13E-02  &8.89E-03 &$\rm ^{52}Cr$ &4.31E-04 &1.39E-03  &2.66E-03 \\
$\rm ^{26}Mg$ &1.85E-03 &1.52E-02  &9.90E-03 &$\rm ^{53}Cr$ &1.03E-04 &1.76E-08  &6.34E-08 \\
$\rm ^{27}Al$ &1.35E-03 &4.31E-03  &2.56E-03 &$\rm ^{54}Cr$ &2.67E-08 &5.77E-11  &5.80E-12 \\
$\rm ^{28}Si$ &9.19E-03 &1.42E-02  &1.33E-02 &$\rm ^{55}Mn$ &1.56E-03 &6.20E-08  &3.70E-07 \\
$\rm ^{29}Si$ &8.32E-03 &9.68E-02  &1.05E-01 &$\rm ^{54}Fe$ &6.83E-04 &3.19E-07  &2.28E-06 \\
$\rm ^{30}Si$ &1.09E-05 &7.40E-03  &1.10E-02 &$\rm ^{56}Fe$ &4.40E-02 &2.33E-01  &2.84E-01 \\
$\rm ^{31}P$  &2.19E-03 &2.97E-02  &3.58E-02 &$\rm ^{57}Fe$ &7.78E-04 &5.37E-05  &8.75E-05 \\
$\rm ^{32}S$  &5.35E-03 &3.97E-02  &3.80E-02 &$\rm ^{59}Co$ &8.86E-06 &2.90E-06  &3.61E-06 \\
$\rm ^{33}S$  &1.72E-03 &6.64E-02  &9.82E-02 &$\rm ^{58}Ni$ &4.83E-04 &4.40E-04  &6.57E-04 \\
$\rm ^{34}S$  &4.31E-06 &4.10E-03  &7.46E-03 &$\rm ^{60}Ni$ &3.07E-04 &2.17E-03  &7.36E-03 \\
$\rm ^{35}Cl$ &4.13E-04 &1.71E-02  &3.40E-02 &$\rm ^{62}Ni$ &1.41E-06 &9.50E-06  &1.67E-04 \\
$\rm ^{37}CL$ &1.34E-04 &1.90E-02  &2.56E-02 &$\rm ^{63}Cu$ &1.78E-08 &4.70E-07  &2.61E-07 \\
$\rm ^{36}Ar$ &1.71E-03 &2.39E-02  &4.63E-02 &$\rm ^{65}Cu$ &3.99E-09 &1.66E-08  &1.02E-06 \\
$\rm ^{38}Ar$ &4.78E-06 &2.28E-03  &3.56E-03 &$\rm ^{64}Zn$ &2.17E-06 &1.28E-05  &8.75E-05 \\
$\rm ^{39}K$  &4.41E-05 &8.04E-03  &1.57E-02 &$\rm ^{66}Zn$ &3.31E-10 &3.67E-09  &8.05E-08 \\
$\rm ^{41}K$  &1.53E-05 &1.36E-02  &1.21E-02 &$\rm ^{56}Ni$ &4.39E-02 &2.32E-01  & 2.84E-01\\
$\rm ^{40}Ca$ &2.54E-03 &5.08E-02  &5.96E-02 &$\rm ^{56}Ni_{\rm Jet}$ &2.65E-02&2.30E-01 &1.95E-01 \\
$\rm ^{44}Ca$ &1.05E-05 &1.71E-03  &2.06E-02 &$\rm ^{56}Ni_{\rm SN}$  &1.75E-02&2.29E-03 &8.82E-02 \\
\hline
\end{tabular*}
\end{center}
\tablenum{3}
\caption{Abundance of heavy elements in the ejecta. 
Abundances are in units of $M_{\odot}$. All unstable nuclei produced
in the ejecta are assumed to decay to the corresponding stable
nuclei. The amount of $\rm ^{56}Ni$ is also in shown in the last row. 
We define the $\rm ^{56}Ni_{\rm Jet}$ as the one within the
$10^{\circ}$ cone around the rotation axis at the final stage of the
calculation, and the left we call as $\rm ^{56}Ni_{\rm SN}$ in this study.
\label{tab2}}
\end{table*}

\end{document}